\documentclass[12pt]{article}

\usepackage{amsmath,amsfonts,amssymb}
\usepackage{graphicx}
\usepackage{geometry}
\usepackage{lmodern} 
\usepackage{caption}
\usepackage{fancyhdr}
\usepackage{lastpage}
\usepackage[backend=biber,style=authoryear,uniquename=false,maxcitenames=2]{biblatex} 
\usepackage{bm}
\usepackage{mathrsfs} 
\usepackage{tikz}
\usepackage{float}
\usepackage{enumitem}
\usepackage{placeins}
\usepackage{pgfplots}
\pgfplotsset{compat=1.18}
\usepackage{booktabs}
\usepackage{lipsum}
\usepackage{graphicx}
\usepackage{xpatch} 
\usepackage[colorlinks=true,allcolors=blue]{hyperref} 
\usepackage[exponent-product=\cdot]{siunitx}
\usepackage{tikz}
\usetikzlibrary{arrows.meta}
\usepackage{mathtools}
\usepackage{subcaption}

\usepackage{listings}
\usepackage{xcolor}
\usepackage{pifont}
\newcommand{\cmark}{\textcolor{green!60!black}{\ding{51}}} 
\newcommand{\xmark}{\textcolor{red!70!black}{\ding{55}}}  

\DeclareCiteCommand{\cite}[\mkbibparens]
  {\boolfalse{citetracker}%
   \boolfalse{pagetracker}%
   \usebibmacro{prenote}}
  {\color{blue}\usebibmacro{citeindex}%
   \usebibmacro{cite}}
  {\multicitedelim}
  {\usebibmacro{postnote}}

\title{\textbf{\LARGE Research and Prototyping Study of an LLM-Based Chatbot for \\ Electromagnetic Simulations}}
\date{}
\author{Albert Piwonski\footnote{Theoretische Elektrotechnik, TU Berlin, Einsteinufer 17, 10587 Berlin, Germany. A. Piwonski can be contacted via e-mail (a.piwonski@tu-berlin.de) or \href{https://www.linkedin.com/in/albert-piwonski-126345222/}{LinkedIn}.} \! and Mirsad Hadžiefendić\footnote{Author names appear in alphabetical order of first name, reflecting equal contributions to the research and manuscript preparation. M. Hadžiefendić can be contacted via \href{https://www.linkedin.com/in/mhadziefendic/}{LinkedIn}.}}

\renewenvironment{abstract}
 {\par\noindent\textbf{\abstractname \\}\ \ignorespaces}
 {\par\medskip}

\usepackage{url}
\usepackage{hyperref}
\usepackage{eso-pic}
\AddToShipoutPicture*{\footnotesize\raisebox{1cm}{\hspace{2.55cm}\fbox{\parbox{\textwidth}{This paper has been submitted to COMPEL for possible publication, published by Emerald Publishing Limited.}}}}

\usepackage[table]{xcolor}
\usepackage{array}
\usepackage{makecell} 
\usepackage{graphicx} 
\usepackage{eurosym} 

\begin{document}

\maketitle
\thispagestyle{empty}


\begin{abstract} \vspace{0.1cm}

\noindent \textbf{Purpose} -- This work addresses the question of how generative artificial intelligence can be used to reduce the time required to set up electromagnetic simulation models. A chatbot based on a large language model is presented, enabling the automated generation of simulation models with various functional enhancements.\vspace{0.1cm}

\noindent \textbf{Design/methodology/approach} -- A chatbot-driven workflow based on the large language model Google Gemini-2.0-Flash automatically generates and solves two-dimensional finite element eddy current models using Gmsh and GetDP. Python is used to coordinate and automate interactions between the workflow components. The study considers conductor geometries with circular cross-sections of variable position and number. Additionally, users can define custom post-processing routines and receive a concise summary of model information and simulation results. Each functional enhancement includes the corresponding architectural modifications and illustrative case studies.\vspace{0.1cm}


\noindent \textbf{Findings} -- With a defined set of functionalities, the chatbot successfully sets up and solves electromagnetic simulation models. Notably, it automatically infers not only Python code but also the domain-specific language code for GetDP. The case studies conducted revealed open research challenges, particularly with regard to the question of how to ensure that results are both syntactically and semantically valid.\vspace{0.1cm} 

\noindent \textbf{Originality/value} -- Currently, the application of machine learning methods to solve electromagnetic boundary value problems is an active area of research (see, e.g., physics-informed neural networks or neural operators). However, to the best of our knowledge, little research has examined the potential of artificial-intelligence-assisted generation of simulation models that prioritizes code generation and execution rather than the enhancement of numerical solution schemes. We leverage a large language model and design tailored workflows that contextualize it through carefully constructed system prompts.\vspace{0.1cm} 

\vspace{0.1cm}

\noindent \textbf{Paper type} -- {Research paper}

\end{abstract}
\newpage


\section{Introduction}
\label{sec:introduction}

The application of machine learning (ML) methods, a subfield of artificial intelligence~(AI), to the solution of electromagnetic boundary value problems (BVPs) is currently a highly active area of research. Deep neural networks for operators~\cite{neural_operators, LuNLoperators, fourierno} and physics-informed neural networks~\cite{raissi1, raissi2, Karniadakis}, in which information about the BVP (and possibly measurement data) is integrated into the loss function of the network, often aim to replace traditional numerical methods such as the finite element (FE) method. For physics-informed neural networks in electromagnetism specifically, see, e.g.,~\cite{pinnmaxwell, Pinn_Guo, pinn_rodrigo}. Furthermore, deep neural networks have also been applied to reduced order modeling; see, e.g.,~\cite{ROM}.

This work addresses an orthogonal problem: How can AI methods be used to reduce the time required to set up electromagnetic simulation models, rather than solving the numerical models themselves? The focus is thus on the assisted generation of simulation models, whereby the numerical scheme itself remains unaffected. A conceptually related direction has recently emerged in the computational fluid dynamics (CFD) community. In~\cite{yue2025foamagentautomatedintelligentcfd}, an AI-based multi-agent framework called Foam-Agent is introduced that supports users in performing simulations based on the open source CFD software \mbox{OpenFOAM}~\cite{open_foam}. To the best of the authors' knowledge, there are only a few works on this subject in computational electromagnetics (see, e.g.,~\cite{CEMagent}).

Motivated by this research gap, the present work introduces a chatbot-driven workflow based on the text capabilities of the multimodal large language model (LLM) Google Gemini-2.0-Flash~\cite{google_gemini_flash_2} that facilitates the automated generation of two-dimensional eddy current simulation models using the open source FE tools Gmsh \cite{Gmsh} and GetDP~\cite{GetDP}. The study focuses on geometries consisting of conductors with circular cross-sections, whose positioning and number can be controlled by the user via natural language prompts. Further, users can define custom post-processing routines, such as a visualization of the ohmic power loss density for a selected subset of conductors. In addition, the user is automatically provided with a textual summary of the model information and simulation results. 

The remainder of this paper is organized as follows: Section~\ref{sec:workflow} introduces the model problem, the main software tools, and the basic workflow of the proposed chatbot prototype. Section~\ref{sec:Architectural_extensions_and_case_studies} discusses architectural extensions of the workflow and case studies that demonstrate domain-specific language (DSL) code inference and automatic textual summarization. In addition, we introduce a possible encoding scheme describing what qualifies as an acceptable result from the user’s point of view. Section~\ref{sec:evaluation} presents an evaluation study quantifying the performance of the chatbot-driven workflow. Different LLM models are benchmarked here on exemplary problems of varying difficulty. Finally, Section~\ref{sec:conclusions_and_outlook} summarizes the main findings and outlines future research directions.


\section{Overview of the basic AI workflow}
\label{sec:workflow}

This section introduces the two-dimensional eddy current model problem and briefly discusses the open source finite element tools Gmsh and GetDP, focusing on their interfacing capabilities. These tools are subsequently integrated into a chatbot-driven workflow, which is presented here in its basic form.


\subsection{Model problem}

\begin{figure}[h!]
    \centering
    \includegraphics[width=0.4\linewidth]{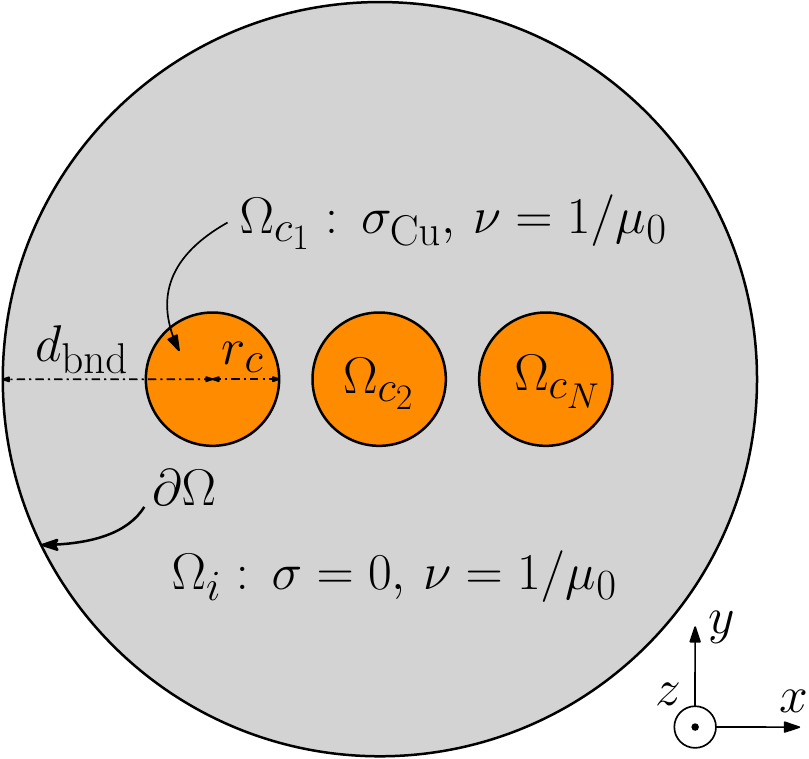}
    \caption{Sketch of the computational domain~$\Omega$: Insulating domain~$\Omega_i$, boundary of the domain~$\partial\Omega$ and $N$ separated conductors, each with a radius~$r_c$. The total conducting domain is $\Omega_c =\cup_{i\in N}\, \Omega_{c_i}$. Each conductor center has at least the minimal distance~$d_{\text{bnd}}$ to the circular boundary~$\partial\Omega$.}
    \label{fig:sketch_bvp_domain}
\end{figure}

As an application example, we consider the translationally symmetric eddy current problem shown in Figure~\ref{fig:sketch_bvp_domain}. Since the electromagnetic fields do not change with respect to the $z$-direction, the problem can be treated in two spatial dimensions. 

We consider $N\in\mathbb{N}$ separated massive conductors with circular cross-sections of radius~$r_c$ and electrical conductivity~$\sigma>0$. The position of the $i$-th conductor is defined by its center point~$\mathbf{p}_{i} \in \mathbb{R}^2$, with $i\in N$. We denote the conducting domain as $\Omega_c=\cup_{i\in N}\, \Omega_{c_i}$, which is surrounded by the insulating domain~$\Omega_i$ with $\sigma = 0$. The total domain \mbox{$\Omega = \Omega_c \cup \Omega_i$} has a circular boundary~$\partial\Omega$.
Since the positions of the conductors are considered variable, it must be ensured that the boundary~$\partial\Omega$ fully encloses all of them, including a minimal distance~$d_{\text{bnd}}$ to reduce the influence of the boundary on the solution. To ensure this, the centroid of all conductors is computed, which defines the center of the circular boundary $\partial\Omega$. The radius of the boundary~$\partial\Omega$ is determined by the distance from the centroid to the outermost conductor center~$\mathbf{p}_i$, plus the constant $d_{\text{bnd}}$, see again Figure~\ref{fig:sketch_bvp_domain}.
We model the boundary~$\partial\Omega$ as perfectly electrically conductive, i.e.,~$\sigma \rightarrow \infty$. Furthermore, magnetic materials are excluded from this application example, so the magnetic reluctivity is $\nu = 1/\mu_0$ in the entire domain.

The model is excited by imposed currents, which define the global conditions of the associated eddy current BVP: Each conductor carries a time-harmonic current of amplitude~$I$ with frequency~$f$, respectively, angular frequency $\omega = 2\pi f$. For the sake of clarity, all parameters of the model problem are summarized in Table~\ref{tab:bvp_data}.

\begin{table}[htbp]
\caption{Parameters of the eddy current model problem}
\centering
\begin{tabular}{l p{6cm} l} 
\toprule
\textbf{Parameter} & \textbf{Description} & \textbf{Numerical value and unit} \\
\midrule
$r_c$ & radius of the conductors & $5\,\text{mm}$ \\
$d_{\text{bnd}}$ & minimum distance between the center of the outermost conductor and the boundary $\partial\Omega$& $3r_c = 15\,\text{mm}$ \\
$\nu$ & magnetic reluctivity & $1/\mu_0$, with $4\pi \times 10^{-7}\,\text{Vs/Am}$ \\
$\sigma$ & electrical conductivity & $0$ in $\Omega_i$, $\sigma_{\mathrm{Cu}} = 58.1\, \text{MS/m}$ in $\Omega_c$\\
$I$ & amplitude of the imposed current & $1\,\text{A}$ \\
$f = 2\pi/\omega$ & frequency of the imposed current & $50\,\text{Hz}$ \\
\bottomrule
\end{tabular}
\label{tab:bvp_data}
\end{table}

\noindent The physical phenomena are governed by the time-harmonic Maxwell's equations in the magnetoquasistatic limit:
\begin{align}
    \mathrm{div}\,\mathbf{B} = 0, \quad \mathbf{curl}\,\mathbf{E } = -j\omega \mathbf{B}, \quad \mathbf{curl}\,\mathbf{H} = \mathbf{J}, \label{eq:governing_equations}
\end{align}
in which $\mathbf{B}$ is the magnetic flux density, $\mathbf{E}$ is the electric field, $\mathbf{H}$ is the magnetic field, $\mathbf{J}$~is the current density, and $j = \sqrt{-1}$ is the imaginary unit. As we consider linear materials, the following constitutive laws hold:
\begin{align}
    \mathbf{B} = \nu \mathbf{H}, \quad \mathbf{J} = \sigma \mathbf{E},\, \text{in}\, \Omega_c,\quad  \mathbf{J} = \mathbf{0}, \,\text{in}\, \Omega_i. \label{eq:constitutive_laws}
\end{align}
The perfectly electrically conductive boundary~$\partial\Omega$ leads to vanishing tangential components of the electric field~$\mathbf{E}$:
\begin{align}
    \mathbf{E} \times \mathbf{n}|_{\partial\Omega} = \mathbf{0}, \label{eq:local_bc}
\end{align}
in which $\mathbf{n}$ is the unit external normal vector of the boundary $\partial\Omega$. \\

Equations~\eqref{eq:governing_equations}, \eqref{eq:constitutive_laws}, \eqref{eq:local_bc}, together with the global conditions for the conductor currents, define an eddy current BVP that can be solved numerically using the finite element method. A commonly used finite element formulation for this type of problem is the modified $\mathbf{A}-v$ magnetic vector potential formulation, see, e.g., ~\cite{j_dular_diss}. Here, the magnetic vector potential~$\mathbf{A}$ satisfying $\mathbf{B} = \mathbf{curl}\,\mathbf{A}$ is introduced in~$\Omega$, while an electric scalar potential~$v$ is defined only in $\Omega_c$. The electric field~$\mathbf{E}$ can be expressed in terms of the scalar and vector potentials as $\mathbf{E} = -j\omega\mathbf{A} -\mathbf{grad}\,v$. To ensure the uniqueness of the magnetic vector potential~$\mathbf{A}$, a gauge condition must be imposed. For the translationally symmetric model the current density $\mathbf{J}$ is assumed to be purely $z$-directed, such that the magnetic flux density~$\mathbf{B}$ only has in-plane components, i.e., $\mathbf{B} = \mathbf{e}_x B_x + \mathbf{e}_y B_y$. This is ensured by choosing~$\mathbf{A} = \mathbf{e}_z A_z$, which implicitly fulfills the Coulomb gauge $\mathrm{div}\,\mathbf{A} = \partial A_z/\partial z = 0$, as the component~$A_z$ solely depends on the coordinates $x$ and $y$.\\ 

Concisely, the $\mathbf{A}-v$ formulation, expressed in its weak (variational) form, can be stated follows: Seek $\mathbf{A} \in \mathcal{A}(\Omega)$ and $\mathbf{grad}\,v \in \mathcal{V}(\Omega_c)$, such that for all test functions $\mathbf{A}' \in \mathcal{A}_0(\Omega)$ and $\mathbf{grad}\,v' \in \mathcal{V}_0(\Omega_c)$:
\begin{align}
    \int_\Omega \nu \, \mathbf{curl}\,\mathbf{A} \cdot \mathbf{curl}\,\mathbf{A}'\, \mathrm{d}\Omega + \int_{\Omega_c}j\omega\sigma\,\mathbf{A} \cdot \mathbf{A}'\,\mathrm{d}\Omega +  \int_{\Omega_c}\sigma\,\mathbf{grad}\,v \cdot \mathbf{A}'\,\mathrm{d}\Omega &= 0, \label{eq:weak_formulation_part_1}\\ 
    \int_{\Omega_c} j\omega\sigma\, \mathbf{A} \cdot \mathbf{grad}\,v' \,\mathrm{d}\Omega + \int_{\Omega_c} \sigma\, \mathbf{grad}\,v \cdot \mathbf{grad}\,v' \,\mathrm{d}\Omega &= \sum_{i\in N} I_i V_i'. \label{eq:weak_formulation_part_2}
\end{align}

The discrete counterpart of the function space~$\mathcal{A}(\Omega)$ is spanned by $z$-directed nodal basis functions that vanish on the boundary, i.e., $A_z|_{\partial\Omega} = 0$. The function space~$\mathcal{V}(\Omega_c)$ is represented by $N$ $z$-directed constants, one associated with each conductor. The term $V_i'$ denotes the voltage that can be associated to the $i$-th conductor. As the current~$I_{i\in N} = I$ is fixed for each conductor, the voltages are resulting from the finite element solution. For additional information, the reader is referred to~\cite{j_dular_diss}.



\subsection{Open source finite element tools Gmsh and GetDP}
\label{subsec:open_source_fe_tools}

The computational domain~$\Omega$ shown in Figure~\ref{fig:sketch_bvp_domain} is discretized using the open source finite element mesh generator Gmsh. Here, the coupling with our later chatbot-driven workflow is achieved via its Python~\cite{python} application programming interface (API). This interface enables the implementation of a Python function that accepts as input a list of tuples of length~2, each specifying the coordinates of a conductor’s center~$\mathbf{p}_i$, with $i\in N$. The number of conductors~$N$ is directly determined by the length of the input list. This procedure enables mesh generation independently of the specific number and spatial arrangement of the individual conductors. The minimal and maximal mesh characteristic sizes are defined based on the radius of the boundary~$\partial\Omega$ and the radius of the conductors~$r_c$ (see Table~\ref{tab:bvp_data}), in order to obtain a manageable number of triangles (for fast prototyping) independently of the number and spatial arrangement of the conductors. Prior to meshing, a check is performed to ensure that the conducting domains do not overlap. Furthermore, for the discretized counterparts of the conductive subdomains~$\Omega_{c_{i\in N}}$ and the boundary~$\partial\Omega$, so-called physical groups need to be defined and stored in the resulting mesh file (``.msh'' file extension), so that common material properties and boundary conditions can be assigned for the subsequent FE solver. 

The $\mathbf{A}-v$ formulation, see~Equations~\eqref{eq:weak_formulation_part_1}, \eqref{eq:weak_formulation_part_2}, is implemented in the  open source finite element solver GetDP. Unlike Gmsh, GetDP cannot be interfaced directly from Python, as no dedicated API is available. Instead, solver files must be written in GetDP’s domain-specific language and stored as ASCII files with the ``.pro'' file extension. These solver files can then be executed via the command-line interface (CLI). After computation, the resulting field plots, e.g., for the magnetic flux density~$\mathbf{B}$ (``.pos'' file extension) can be visualized using the post-processing facilities provided by Gmsh. It should be emphasized that the CLI commands are executed from Python through the subprocess module.


\subsection{Chatbot-driven workflow}
\label{subsec:basic_workflow}

\begin{figure}[!ht]
    \centering
    \includegraphics[width=1\textwidth]{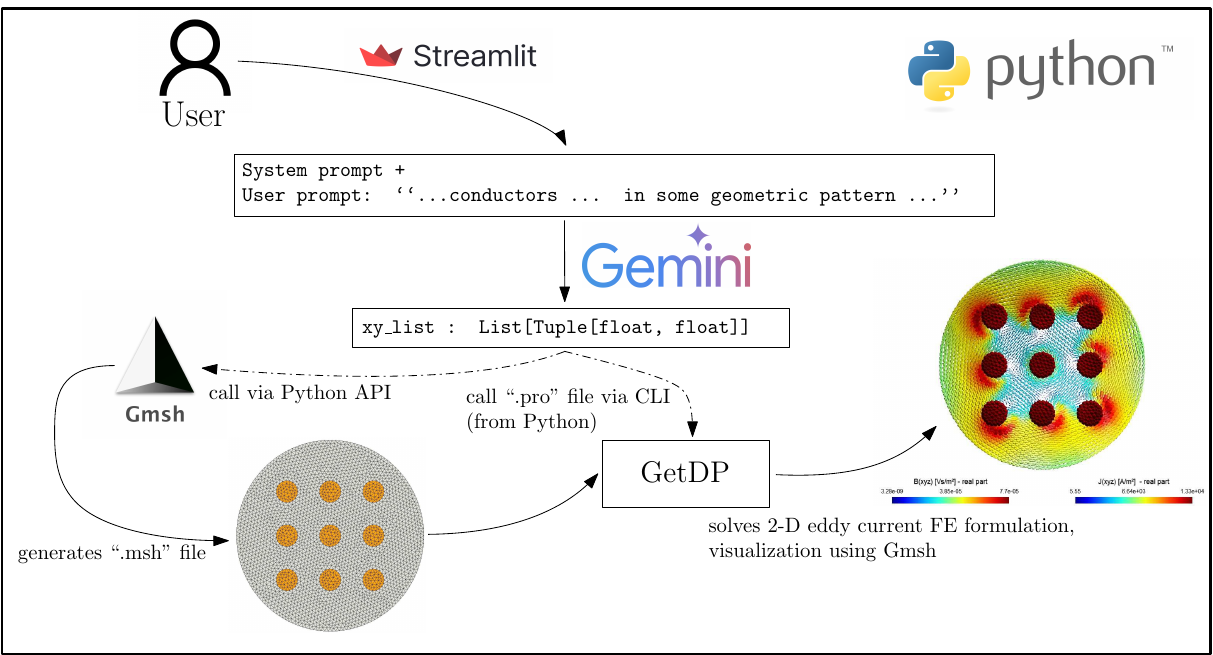}
    \caption{Sketch of the basic AI workflow.}
    \label{fig:basic_workflow}
\end{figure}

\begin{figure}[!h]
    \centering
    \includegraphics[width=.7\textwidth]{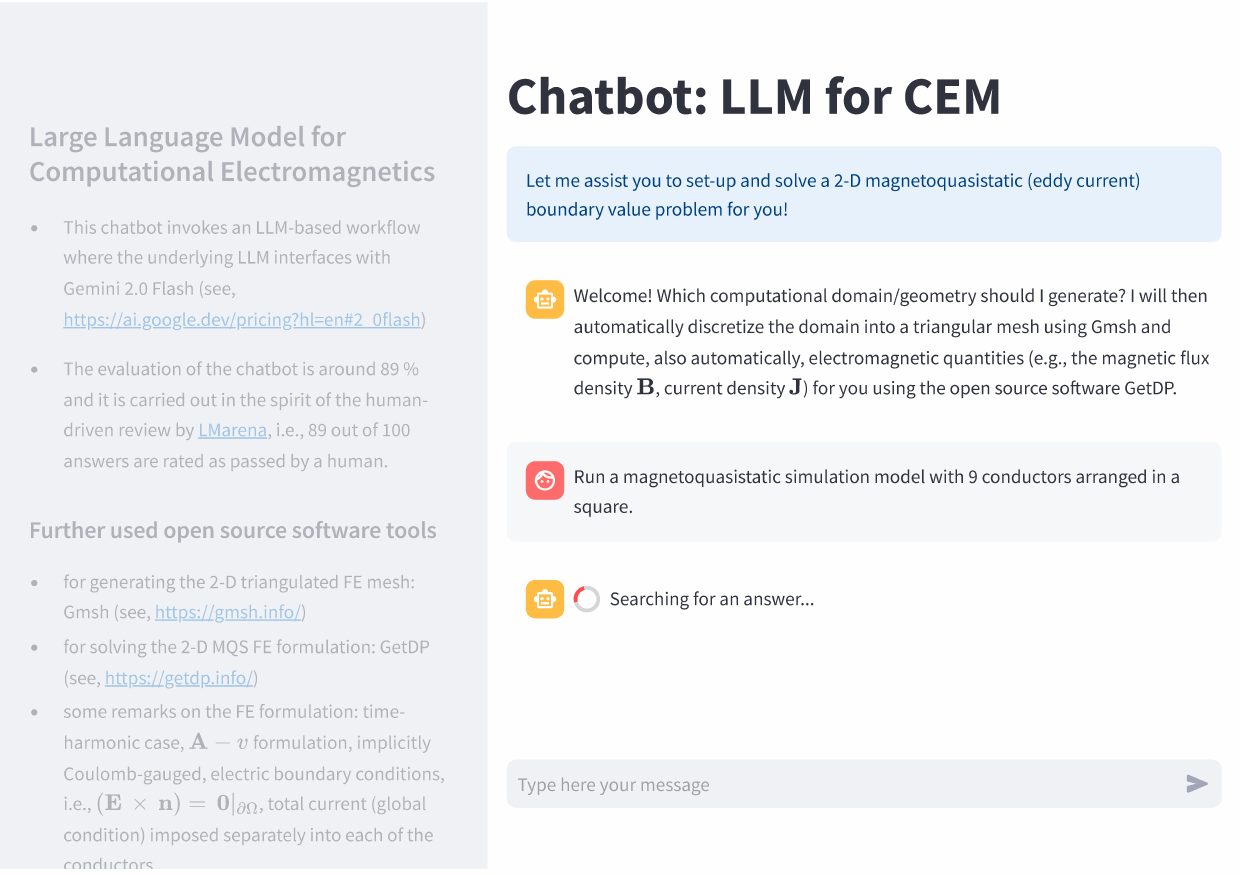}
    \caption{User interface of the developed chatbot: Status bar containing additional meta information (left), initial information~(top) and text box for user prompts (bottom).}
    \label{fig:user_interface}
\end{figure}

Both open source FE tools are now integrated into a chatbot-driven workflow, as illustrated in Figure~\ref{fig:basic_workflow}. The chatbot is based on the free tier of Google Gemini-2.0-Flash, a multimodal LLM, but uses only its text generation capabilities. For general literature on LLMs, see~\cite{LLMbook}. The user interface is implemented as an interactive web application using the open source framework Streamlit~\cite{streamlit2025}, see Figure~\ref{fig:user_interface}. Python serves as the coordinating and automation layer, managing the interaction between workflow components. 

The program workflow can be summarized as follows: The user’s prompt is incorporated into a system prompt containing a task description, rules and examples (see Appendix~\ref{subsec:system_prompts_Python}), which is given to the LLM to generate a string of (ideally\footnote{Mind that the LLM is a probabilistic model where, even though, input and output guardrails are used, there is no complete guarantee that the expected correct structured output is generated and that this output is fully reproducible in a deterministic sense. Thus, even when given the same input, a degree of uncertainty regarding the resulting output remains.}) syntactically correct Python code. Then, a cleaned version of the LLM’s output string serves as the input for a function that executes the dynamically generated Python code. The Gmsh and GetDP code is static and accessed via a predefined Python wrapper function (see Section~\ref{subsec:open_source_fe_tools}), which takes a list of coordinate tuples as input and runs the finite element simulation as a side effect. It is important to emphasize that the LLM’s weights are not updated or re-trained (that is, there is no fine-tuning); rather, the pre-trained model is contextualized at runtime via the system prompt. More precisely, the system prompt supplies task-specific instructions and context regarding the model's input, however, it does not alter the model's parameters. Moreover, this workflow does not employ retrieval-augmented generation (RAG) or any persistent memory mechanism: no external knowledge store is queried during generation and no state is retained beyond the immediate prompt.
 
\vspace{-0.4cm}
\vspace{-0.8cm}


\clearpage
\section{Architectural extensions and case Studies}
\label{sec:Architectural_extensions_and_case_studies}

This Section presents functional enhancements to the basic AI workflow described in Section~\ref{subsec:basic_workflow}. The architectural modifications are detailed, and case studies are conducted and discussed for each autonomy level of the chatbot.


\subsection{Inferring Python code to generate a list of coordinate tuples}
\label{subsec:inf_pyt_code_gen_list_coord_tuples}

\begin{figure}[!ht]
  \centering
  \begin{minipage}[b]{0.48\linewidth}
    \centering
    \includegraphics[width=\linewidth]{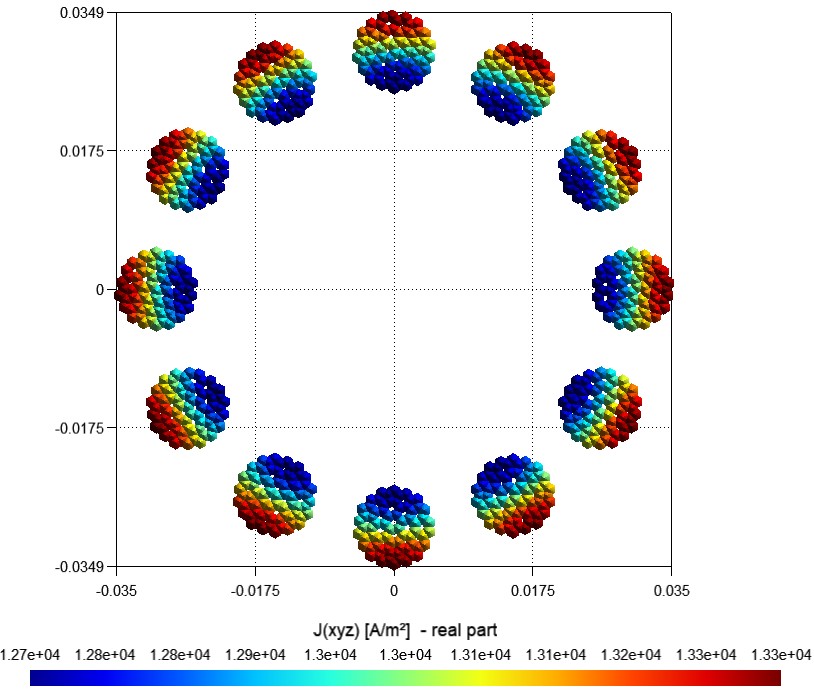}
    \par\smallskip
    \centering\footnotesize (a) Run an eddy current simulation model using~12 conductors following the pattern of a circle with radius $r = 0.03\,\text{m}$.
    \label{fig:2x2-a1}
  \end{minipage}\hfill
  \begin{minipage}[b]{0.48\linewidth}
    \centering
    \includegraphics[width=\linewidth]{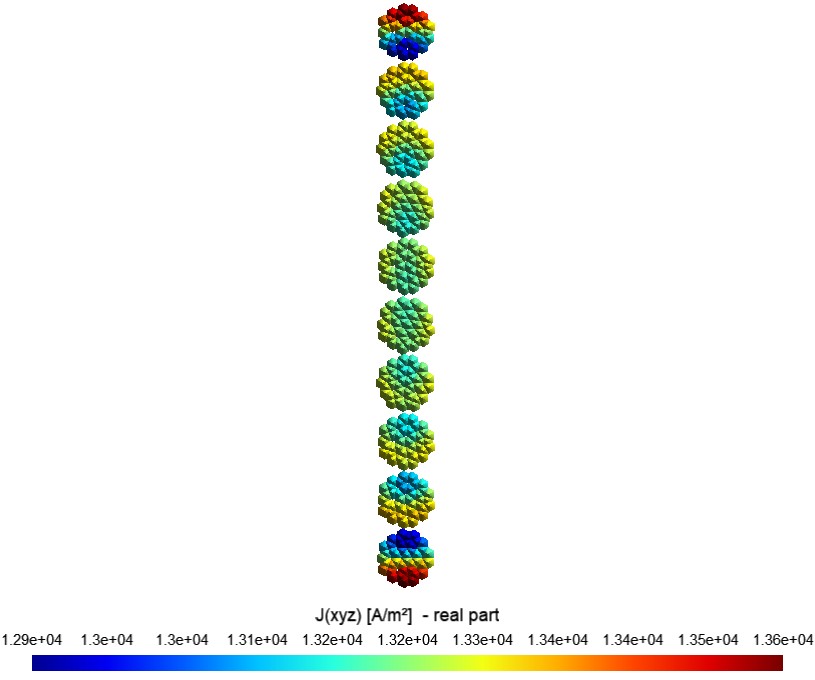}
    \par\smallskip
    \centering\footnotesize (b) Run an MQS simulation with 10~conductors following the $y$-axis. The conductors have a radius of $5\,\text{mm}$; ensure a minimal spacing between them.
    \label{fig:2x2-b1}
  \end{minipage}

  \vspace{6pt} 

  \begin{minipage}[t]{0.48\linewidth}
    \centering
    \includegraphics[width=\linewidth]{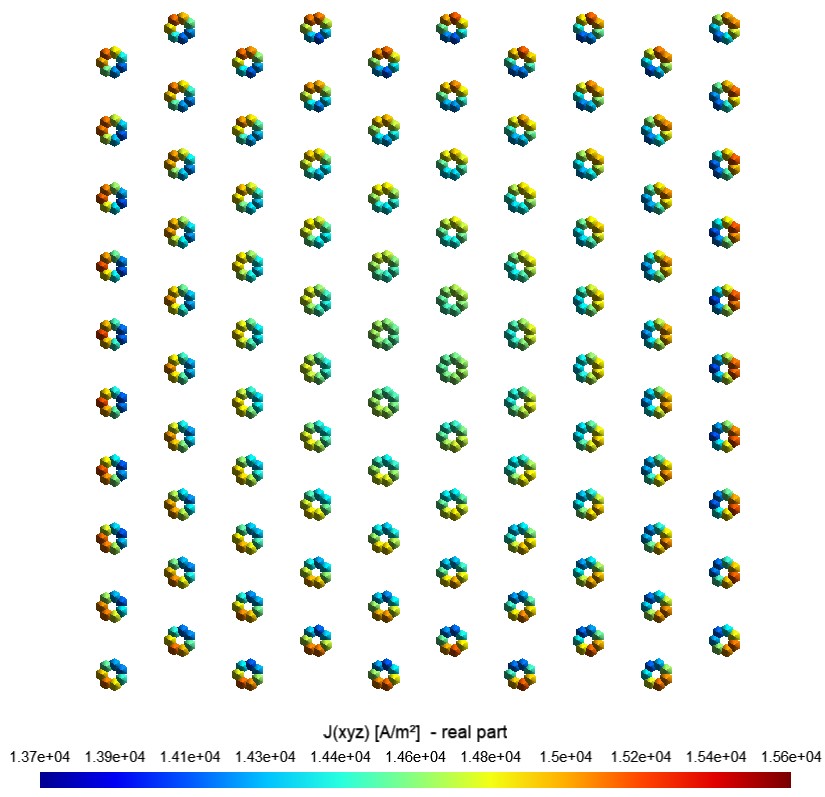}
    \par\smallskip
    \centering\footnotesize (c) Run an MQS simulation with 100 conductors arranged in a $10 \times 10$ grid using hexagonal packing.
    \label{fig:2x2-c1}
  \end{minipage}\hfill
  \begin{minipage}[t]{0.48\linewidth}
    \centering
    \includegraphics[width=\linewidth]{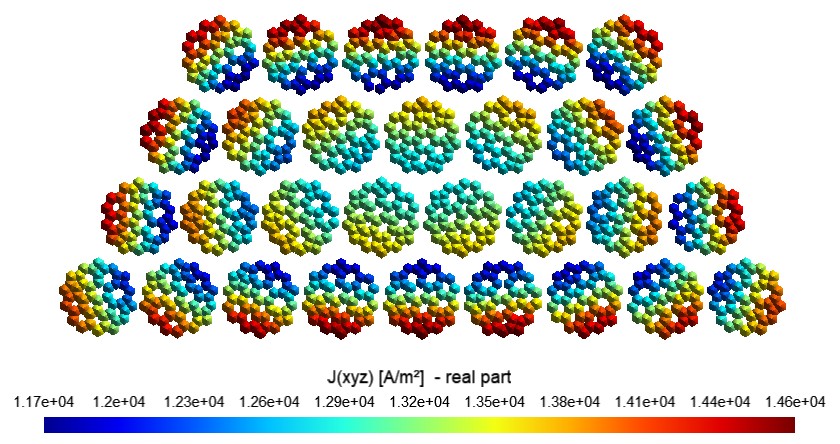}
    \par\smallskip
    \centering\footnotesize (d) Run an MQS simulation model with a proper number of conductors, that fill a trapezoidal slot form. Note, that each conductor has a radius of $5\,\text{mm}$ and must not overlap.
    \label{fig:2x2-d1}
  \end{minipage}

\caption{Real parts of the current density~$\mathbf{J}$ distributions corresponding to the models generated based on user prompts (a)--(d).}
\label{fig:results_inferring_python_code_1}
\end{figure}

The workflow of Section~\ref{subsec:basic_workflow} is used without further modifications to generate the exemplary results shown in Figure~\ref{fig:results_inferring_python_code_1} and~\ref{fig:results_inferring_python_code_2}. Examples of the inferred Python code are provided in Appendix~\ref{subsec:llm_outputs_inferred_python_code}.

\vspace{-0.4cm}

\vspace{-0.5cm}
\begin{figure}[!ht]
  \centering
  \begin{minipage}[b]{0.48\linewidth}
    \centering
    \includegraphics[width=\linewidth]{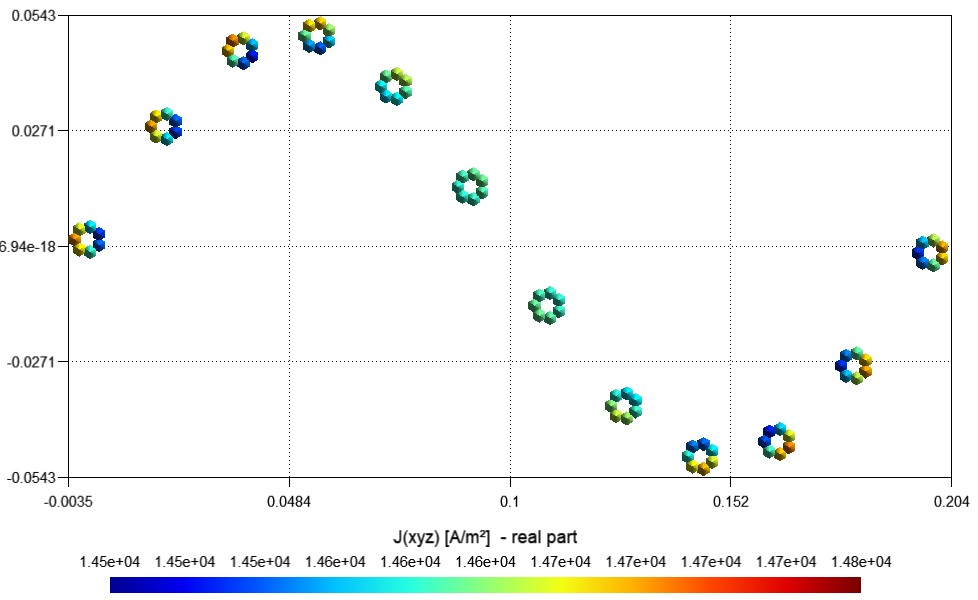}
    \par\smallskip
    \centering\footnotesize (e) Run an MQS simulation with 12 conductors following the curve $y = f(x) = 0.05 \, \sin(\alpha x)$. Scale $\alpha$ in such a way, that at least one spatial sinus period is depicted. Ensure that \mbox{$0\, \text{m} \le x  \le 0.2\,\text{m}$}.
    \label{fig:2x2-a2}
  \end{minipage}\hfill
  \begin{minipage}[b]{0.48\linewidth}
    \centering
    \includegraphics[width=\linewidth]{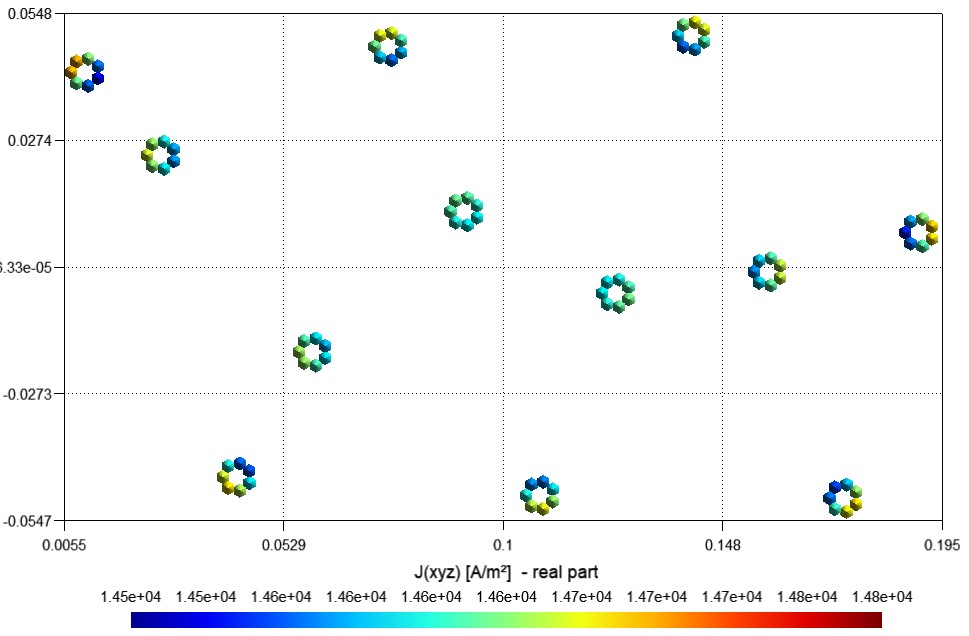}
    \par\smallskip
    \centering\footnotesize (f) Run an MQS simulation with 12 conductors following the curve $y = f(x) = 0.05 \, \sin(\alpha x)$. Scale $\alpha$ in such a way, that at least one spatial sinus period is depicted. Ensure that \mbox{$0\, \text{m} \le x  \le 0.2\,\text{m}$}.
    \label{fig:2x2-b2}
  \end{minipage}

  \vspace{6pt} 

  \begin{minipage}[t]{0.48\linewidth}
    \centering
    \includegraphics[width=\linewidth]{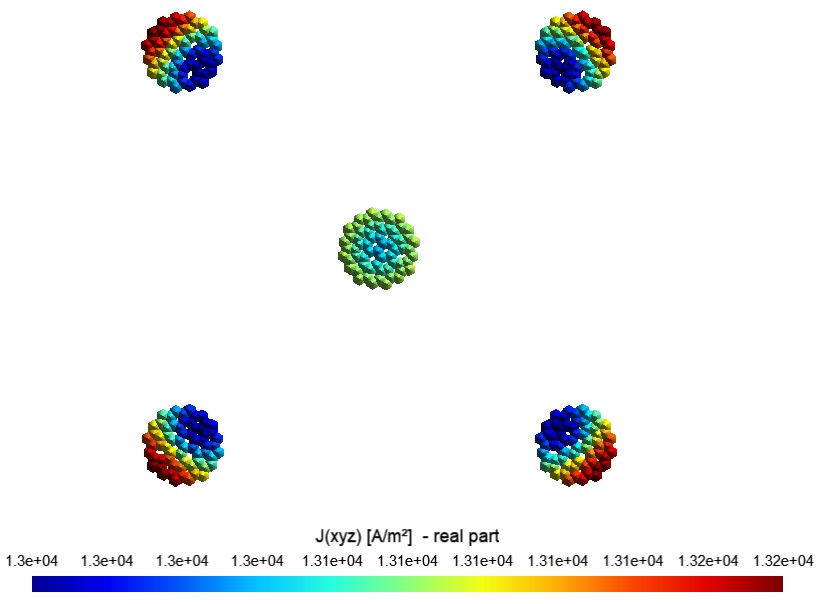}
    \par\smallskip
    \centering\footnotesize (g) Run an MQS simulation model using 5~conductors placed exactly at all 5 vertices of a square with side length $0.05\,\text{m}$.
    \label{fig:2x2-c2}
  \end{minipage}\hfill
  \begin{minipage}[t]{0.48\linewidth}
    \centering
    \includegraphics[width=0.8\linewidth]{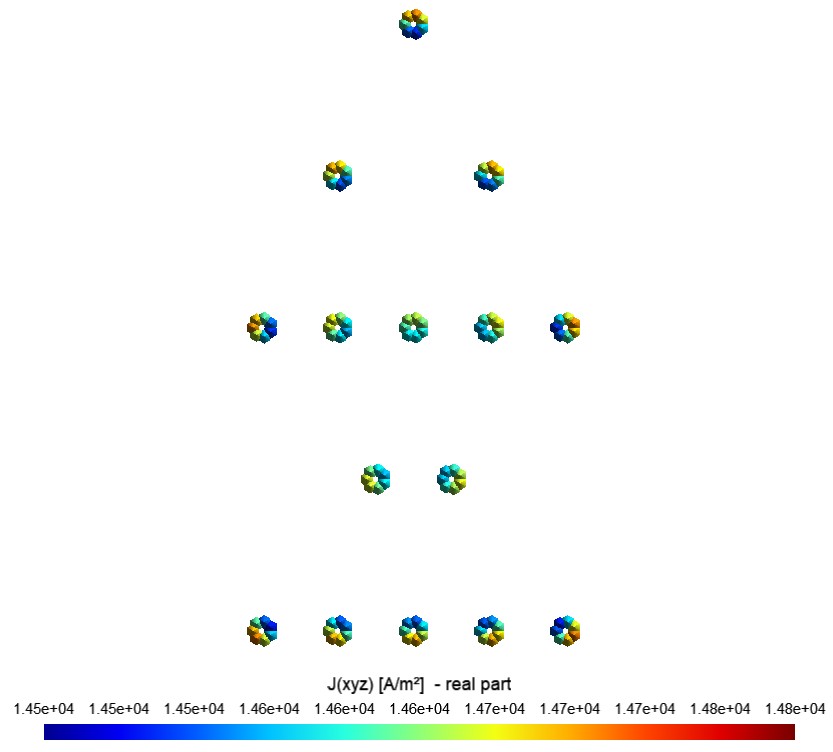}
    \par\smallskip
    \centering\footnotesize (h) Run an MQS simulation using 15 conductors forming the outline of the letter ``A''.
    \label{fig:2x2-d2}
  \end{minipage}
\caption{Real parts of the current density~$\mathbf{J}$ distributions corresponding to the models generated based on user prompts (e)--(h).}
  \label{fig:results_inferring_python_code_2}
\end{figure}

It can be observed that the workflow successfully infers Python code capable of generating conductor arrangements across multiple levels of complexity. The risk of conductor overlap can be significantly reduced if the user’s prompt includes the specific value of the conductor radius~$r_c$, a parameter that is not known a priori to the LLM. The simulation models shown in Figure~\ref{fig:results_inferring_python_code_2}, based on identical user prompts (e) and (f), illustrate the stochastic nature of the underlying LLM. Both generated geometries are correct; however, the Python code for (e) assumes~$\alpha=1$, whereas the code for (f) uses $\alpha = 100$. The resulting simulation model for the user prompt~(g) can be interpreted as hallucination of the LLM, that is, an output inconsistent with reality. However, such hallucinations were observed only rarely in this use case; in this instance, the effect was triggered by the inconsistent user prompt, as a square obviously has only four vertices. From the standpoint of automated model validation, a notable challenge occurs when the inferred Python code is syntactically valid, yet the resulting model exhibits an incorrect geometric interpretation (see the generated model for the user prompt~(h) in Figure~\ref{fig:results_inferring_python_code_2}).


\subsection{Inferring domain-specific language code (including meaningful examples in the system prompt)}
\label{subsec:inf_dsl_code_w_examples}

\begin{figure}[h!]
    \centering
    \includegraphics[width=\linewidth]{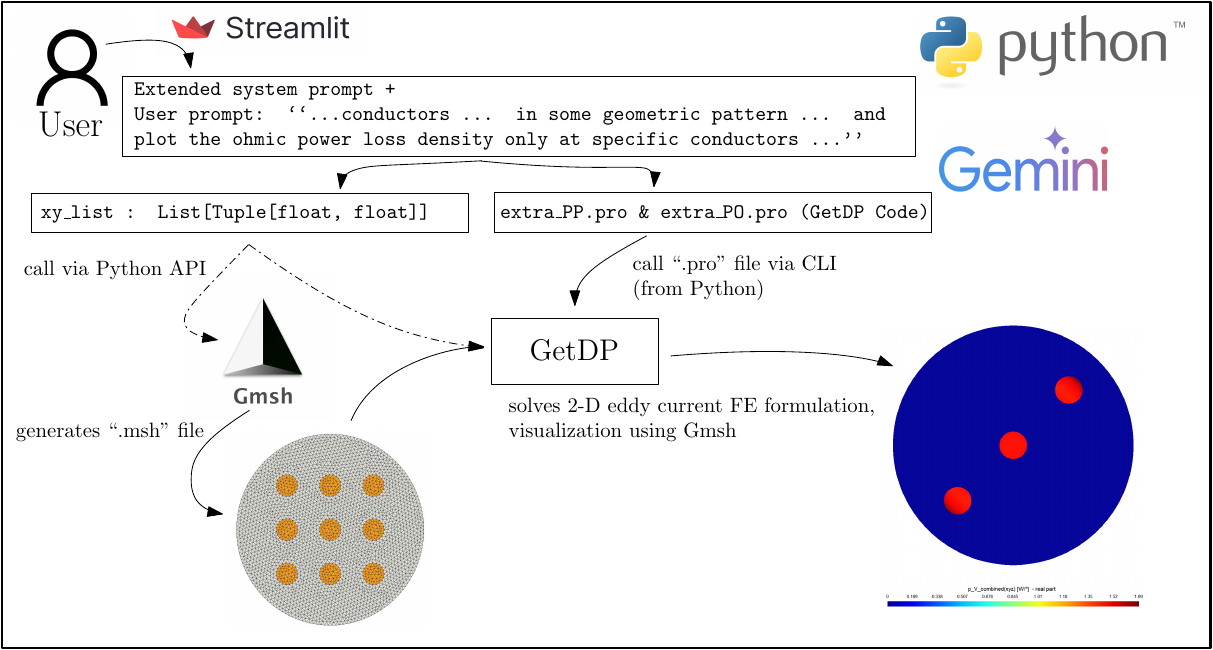}
    \caption{Sketch of the extended AI workflow for dynamically generated post-processing routines (including meaningful examples in the system prompt).}
    \label{fig:workflow_inferring_dsl_code_with_examples}
\end{figure}

The workflow described in Section~\ref{subsec:basic_workflow} is extended to allow the user to define custom post-processing routines; see Figure~\ref{fig:workflow_inferring_dsl_code_with_examples}. As an application example, the visualization of the ohmic power loss density~$p_\Omega$ is considered, which is defined as a time-averaged quantity as follows:
\begin{align}
    p_\Omega = \frac{\mathbf{E} \cdot \mathbf{J}^{*}}{2} = \frac{\sigma|\mathbf{E}|^2}{2} = \frac{\sigma}{2}|-j\omega\mathbf{A} - \mathbf{grad}\,v|^2. \label{eq:ohmic_power_loss_density}
\end{align}

An extended system prompt is provided to the LLM, that contains GetDP code examples for a post-processing routine, that computes and plots the ohmic power loss density~$p_\Omega$ only for a selected subset of conductors, see Appendix~\ref{subsec:system_prompts_including_examples}. Internally, GetDP uses PostProcessing objects to define quantities based on the primary variables of the finite element formulation -- here, the vector potential $\mathbf{A}$ and the gradient of the electric scalar potential~$\mathbf{grad}\,v$. Their visualization and output are controlled via PostOperation objects, which specify formats such as line evaluations, surface plots, or data tables. The dynamically generated GetDP code is written into dedicated ``.pro'' files, which are then included in the main solver file. It should be emphasized that the LLM now infers code in the domain-specific language of GetDP, a language on which, most likely, Google Gemini-2.0-Flash has been trained considerably less such that the LLM's internal knowledge (cf.~\cite[p.~301]{Huyen2025}) about GetDP is most likely quite low.

The extended workflow leads to the exemplary result shown in Figure~\ref{fig:results_inferring_dsl_code}. Further examples of the LLM outputs are provided in the Appendix~\ref{subsec:llm_outputs_inferred_dsl_including_examples}. The results indicate that the basic workflow from the previous subsection can be extended with dynamically generated domain-specific language code. It was observed that including an adequate number of representative code examples significantly reduces syntax errors related to the GetDP language. In their absence, numerous syntax errors occurred, primarily due to missing or superfluous curly brackets.

\begin{figure}[h!]
    \begin{minipage}[t]{0.48\linewidth}
    \centering
    \includegraphics[width=\linewidth]{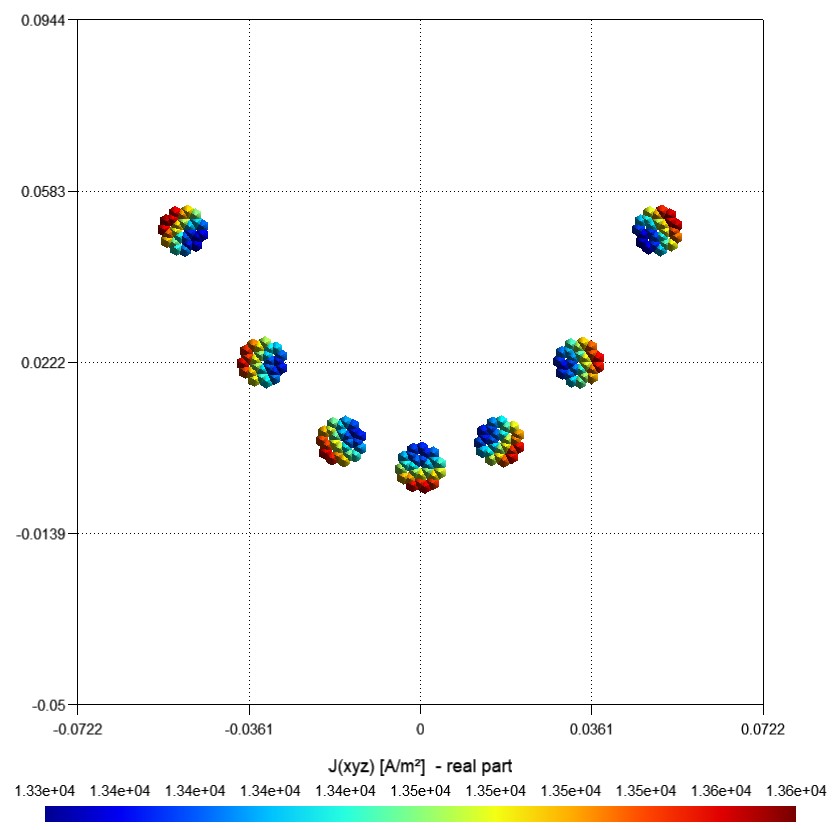}
    \par\smallskip
    \centering\footnotesize (i) Run an MQS simulation using 7 conductors following the parametrization $y(x) = 20x^2$, for $x \in [-0.05, 0.05] \,\text{m}$. Plot the ohmic power loss density only for the conductor whose center point satisfies $f'(x) = 0$.
    \label{fig:2x2-c3}
  \end{minipage}\hfill
  \begin{minipage}[t]{0.48\linewidth}
    \centering
    \includegraphics[width=\linewidth]{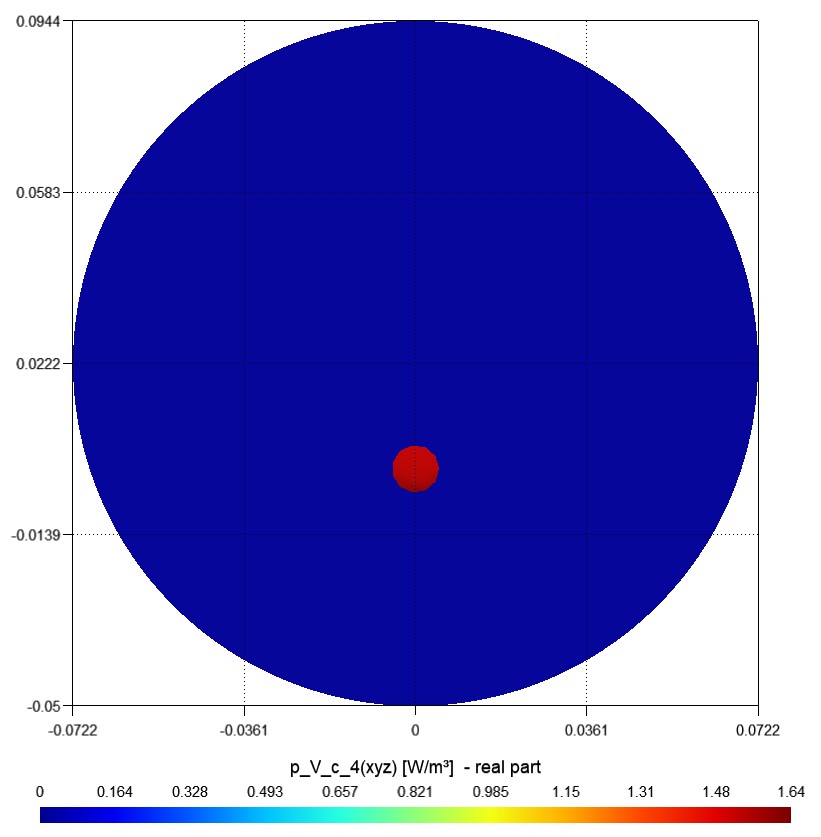}
    \par\smallskip
    \centering\footnotesize Plot of the ohmic power loss density~$p_\Omega$ for the subset of conductors specified in user prompt~(i), i.e., the corresponding GetDP code was correctly dynamically generated and executed.
    \label{fig:2x2-d3}
  \end{minipage}
\caption{Left: Real part of the current density $\mathbf{J}$ corresponding to the simulation model resulting from user prompt~(i), right: Customized post-processing routine, see user prompt~(i).}
  \label{fig:results_inferring_dsl_code}
\end{figure}


\subsection{Inferring domain-specific language code (excluding meaningful examples in the system prompt)}
\label{subsec:inferring_dsl_excluding_examples}

\begin{figure}[ht!]
    \centering
    \includegraphics[width=\linewidth]{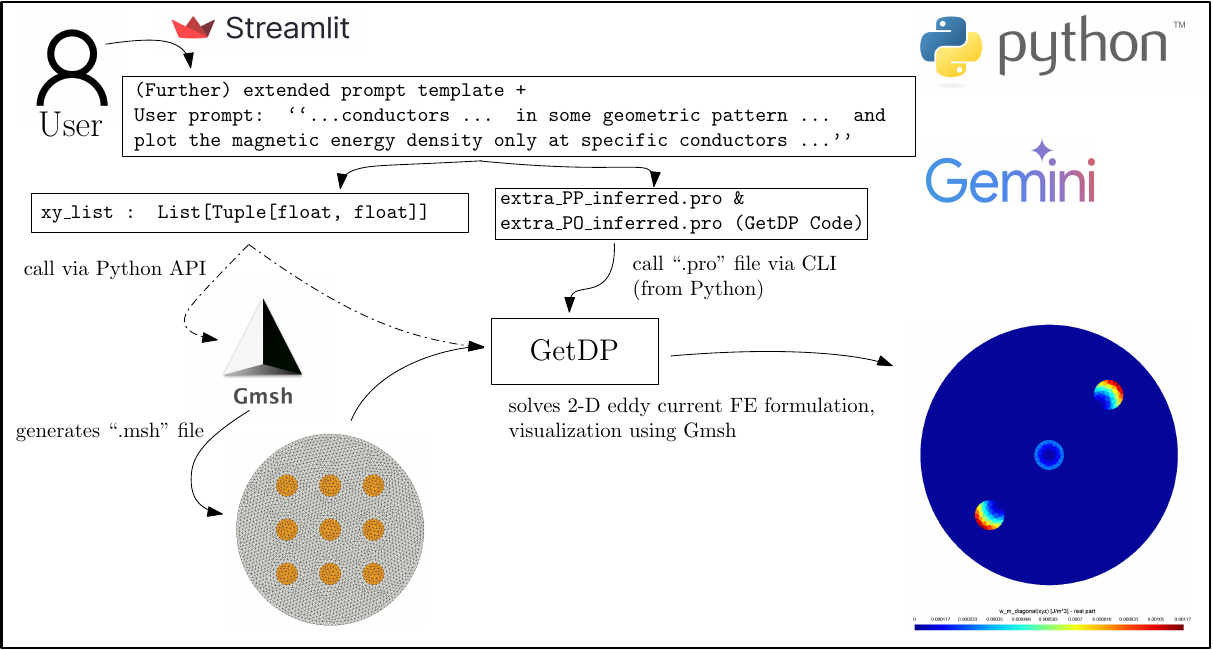}
    \caption{Sketch of the AI workflow for dynamically generated post-processing routines (excluding meaningful examples in the system prompt).}
    \label{fig:workflow_inferring_dsl_code_without_examples}
\end{figure}

In Figure~\ref{fig:workflow_inferring_dsl_code_without_examples}, the main difference to the previous architecture presented in Figure~\ref{fig:workflow_inferring_dsl_code_with_examples} is that (ideally) syntactically and semantically correct domain-specific language code is generated without having any relevant examples in the system prompt. The goal of this architectural modification is to enable valid responses to user prompts such as 
\begin{quote}
“Using nine conductors that are positioned along a rectangle such that one point is at the center (intersection of the diagonals) and the other eight are distributed along the edges of the rectangle, evaluate the plot of the magnetic energy density in terms of the magnetic vector potential vector field within the frequency domain only for the central conductor and the conductors on the first and third bisectors of the rectangle.”
\end{quote}
An exemplary system prompt can be found in the Appendix~\ref{subsec:system_prompts_excluding_examples}.

\newpage
\noindent The distinctive property of this kind of user prompts is that, at the same time, 
\begin{enumerate}[label=(\Roman*)]
      \item these user prompts build upon examples within the system prompt discussed in previous posts such as “only for the central conductor” (cf.~Section~\ref{subsec:inf_dsl_code_w_examples}) and “along a rectangle” (cf.~Section~\ref{subsec:inf_pyt_code_gen_list_coord_tuples}), and
    \item these user prompts refer to examples that are not within the system prompt such as “the magnetic energy density”.
\end{enumerate} 

\noindent Considering the visualization of the magnetic energy density~$w_m$, recall its definition as a time-averaged quantity as follows:
\begin{align}
    w_m = \frac{\mathbf{H} \cdot \mathbf{B}^{*}}{4} = \frac{\nu}{4}|\mathbf{B}|^2 = \frac{\nu}{4}|\mathbf{curl}\,\mathbf{A}|^2. \label{eq:magnetic_energy_density}
\end{align}

A meaningful user prompt provides the LLM-based chatbot with sufficient knowledge regarding (II) such that syntactically and semantically correct DSL code can be inferred. For example, the above-mentioned user prompt has to be extended in such a way such that a contextual answer is received that conforms with the user’s expectation. More precisely, the following sentence has to be added to the above-mentioned user prompt: 
\begin{quote}
``Mind that the following three constraints regarding the magnetic energy density formula should be satisfied: (1) the correct factor 0.25 is present; (2) the material relation between the magnetic flux density vector field and the magnetic vector field holds to be true w.r.t. the reciprocal permeability, i.e., nu; (3) the expression is simplified by using norms.''
\end{quote}

Some observations concerning the inferring of domain-specific language code without using meaningful examples in the system prompt within the chatbot-driven workflow are:

\begin{itemize}
    \item The lack of the extension for the above-mentioned user prompt may result in:
    \begin{itemize}
           \item[-] syntactically incorrect code within the context of the DSL; 
           \item[-] syntactically correct code that is semantically incorrect within the context of the underlying physics (e.g., using a factor of 0.5 vs. 0.25); 
           \item[-] syntactically correct code that behaves semantically correct within the context of the underlying physics, though, semantically incorrect within the context of the DSL (e.g., using the magnetic flux density instead of the magnetic vector potential as primary numerical solution quantity).
    \end{itemize}
    \item Given the system design at hand, there are still many optimization paths regarding the overall prompt engineering in order to, e.g., balance the provision of knowledge by user prompts and system prompts. An interesting approach could be the use of a curated prompt store in a similar manner to a feature store (see, e.g.,~\cite[pp.~325ff]{Huyen2022}) in other machine learning systems.
\end{itemize}


\subsection{Inferring a textual summary of the AI workflow's output}
\label{subsec:textual_summary}

\begin{figure}[h!]
    \centering
    \includegraphics[width=1\linewidth]{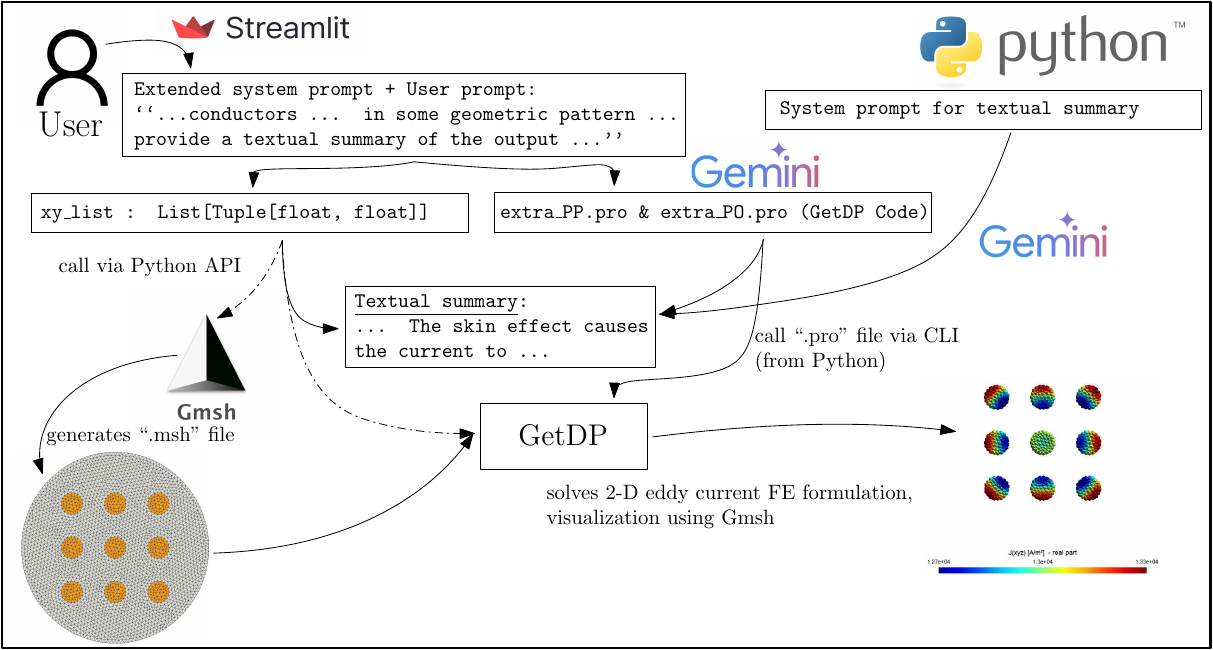}
    \caption{Sketch of the AI workflow for dynamically generated textual summaries.}
    \label{fig:workflow_inferring_textual_summary}
\end{figure}

The main extension of the previous section's workflow is the following: A function is added that executes a second LLM API-call. Its input is the output of the function that executes the first LLM API-call; and its output is a textual summary, see Figure~\ref{fig:workflow_inferring_textual_summary}. Note that there is a system prompt that contains meaningful examples and rules for structuring the output format of the textual summary, see Appendix~\ref{subsec:system_prompts_textual_summary}. In other words, the dynamically generated syntactically and semantically (ideally) correct code is translated into a textual summary that expresses the physical meaning of the code in a natural language.

\begin{figure}[h!]
    \begin{minipage}[b]{0.48\linewidth}
    \centering
    \includegraphics[width=\linewidth]{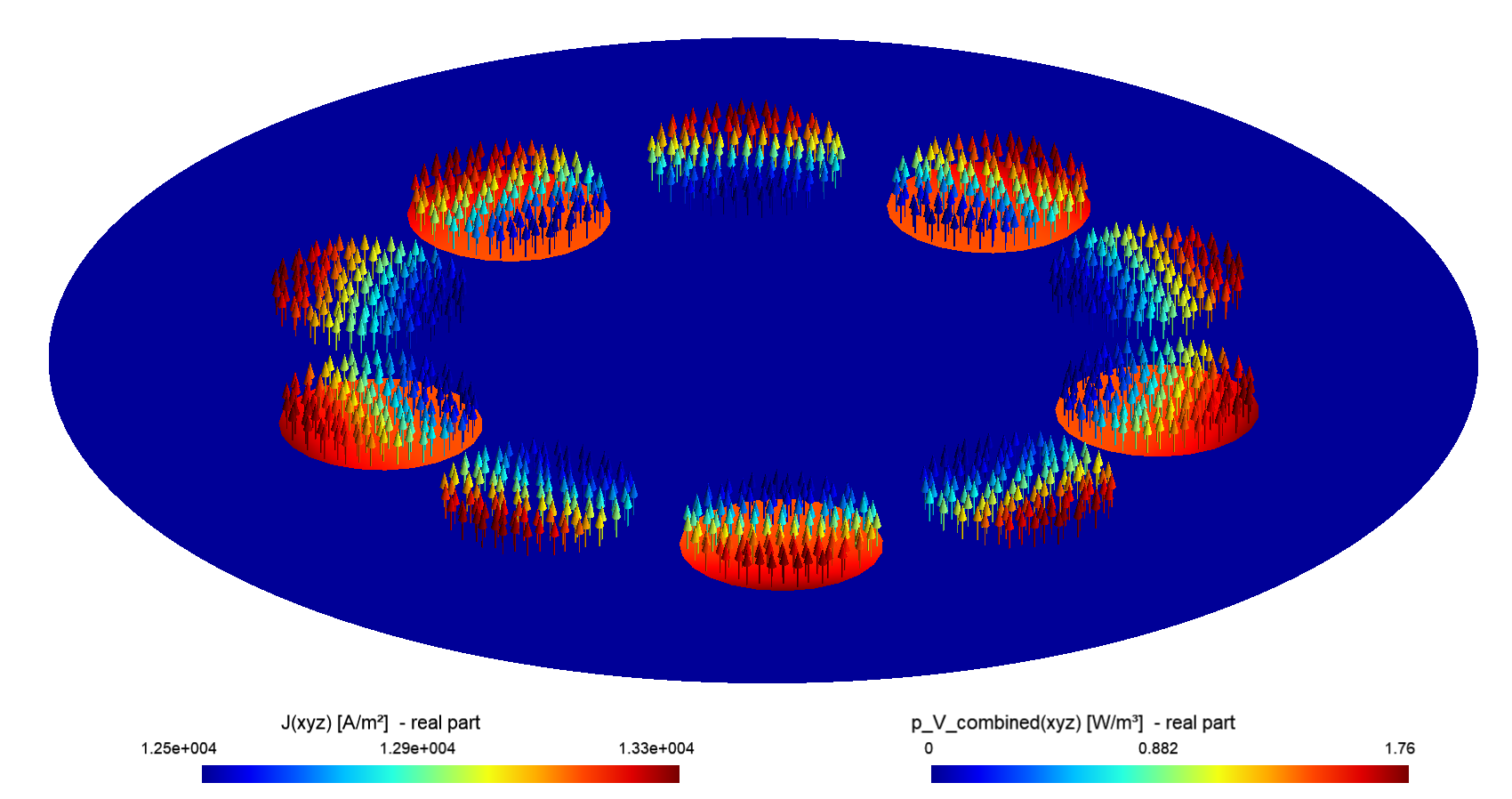}
    \par\smallskip
    \centering\footnotesize (j) Run an MQS simulation using 10 conductors that follow the pattern of a circle. Plot the ohmic loss density only at every second conductor. Provide a summary of the output.
    \label{fig:2x2-c4}
  \end{minipage}\hfill
  \begin{minipage}[b]{0.48\linewidth}
    \footnotesize The plot shows a 2D simulation of 10 conductors arranged in a circle. The ohmic loss density, representing power dissipation due to current flow, is displayed for every second conductor (conductors 1, 3, 5, 7, and 9). The positions of the conductors' center points are determined by a circle of radius 0.02, with coordinates calculated using sine and cosine functions, reflecting their circular arrangement. Due to the alternating current, both skin and proximity effects are present, leading to non-uniform current density distributions within the conductors and influencing the ohmic loss density. The skin effect causes current to concentrate near the surface of each conductor, while the proximity effect modifies the current distribution due to the presence of neighboring conductors.
    \label{fig:2x2-d4}
  \end{minipage}
\caption{Left: Real part of the current density $\mathbf{J}$ and custom post-processing routine for the ohmic power loss density~$p_\Omega$ generated based on user prompt~(j). Right: Dynamically generated textual summary of the results.}
  \label{fig:results_textual_summary}
\end{figure}

A couple of observations and conceptual challenges regarding the textual summary within the chatbot-driven workflow are:

\begin{itemize}
    \item Specifically, the description regarding the skin- and the proximity-effect is reasonable to a certain degree. Since a well-defined metric for quantifying potential improvements in the quality of a textual summary is currently lacking, it remains unclear whether incorporating the skin depth as an additional physical entity in the code would yield a more nuanced textual summary.
    \item Generally, the translation of code into natural language seems to work well to a reasonable extent, see Figure~\ref{fig:results_textual_summary}. Due to the above-mentioned lack of a quality metric for a textual summary, though, it remains unclear whether translating numerical simulation plots (i.e., images) into text or translating numerical simulation numbers (i.e., tabular data) into text could lead to a more nuanced textual summary.
\end{itemize}

\noindent Notice that the conceptual challenge with regard to a quality metric for a textual summary is linked to the fundamental conceptual challenge of how to enable an automated LLM-workflow evaluation.

\subsection{Aspects of semantic and syntactic sources of potential failure in the AI workflow}
\label{subsec:syntax_and_semantics}

The workflows presented in Sections~\ref{subsec:inf_pyt_code_gen_list_coord_tuples}--\ref{subsec:textual_summary} introduce different levels of capabilities in the generated code. Here, we introduce a possible encoding scheme that describes what qualifies as an acceptable result from the user’s point of view.\\

Considering the workflow described in Section~\ref{subsec:inf_pyt_code_gen_list_coord_tuples}, the dynamically generated Python code must first satisfy the syntax and semantics of the Python programming language. Theoretically, all combinations of the truth values listed in Table~\ref{tab:python_true_false} could occur. Syntax errors can arise, for instance, if brackets are left unclosed, as in ``\texttt{(a + 1}'', or if reserved keywords are used as variable names. Semantic errors can occur, for example, when a variable is used before it is defined, or when operations are incompatible with the types of their inputs, as in ``\texttt{'a' + 3}''. While syntactic errors are detected during parsing, both syntax and semantic errors are ultimately revealed when the generated code is executed.

\begin{table}[h!]
\caption{Boolean values representing the syntax and semantics of the dynamically generated Python code.}
\centering
\begin{tabular}{c c} 
\toprule
\textbf{Python syntax} & \textbf{Python semantics}\\
\midrule
\xmark & \xmark\\
\xmark & \cmark\\
\cmark & \xmark\\
\cmark & \cmark\\
\bottomrule
\end{tabular}
\label{tab:python_true_false}
\end{table}

\begin{figure}[h!]
    \centering
    \includegraphics[width=0.5\textwidth]{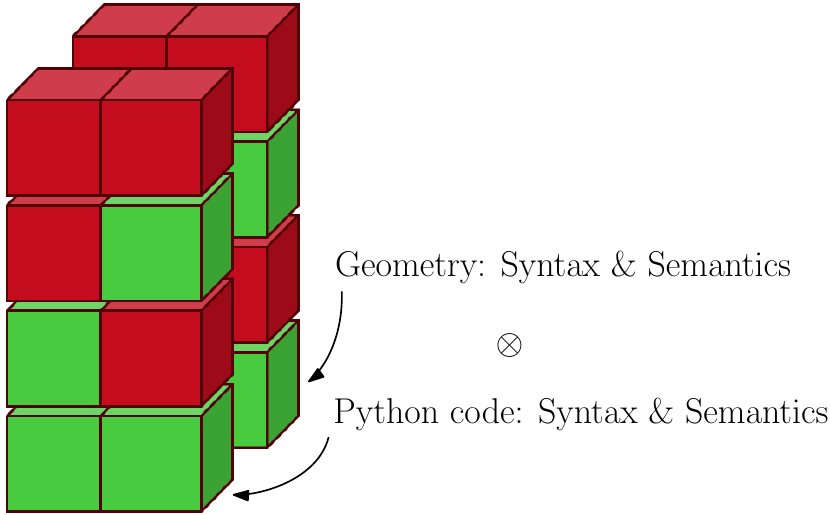}
    \caption{Stack of syntaxes and semantics for workflow as in Section~\ref{subsec:inf_pyt_code_gen_list_coord_tuples}.}
    \label{fig:python_times_geometry}
\end{figure}

More subtle errors occur when the generated Python code is both syntactically and semantically correct, but the resulting geometry does not match the user’s expectations, consider, e.g., the resulting model for the user prompt~(h) in Figure~\ref{fig:results_inferring_python_code_2}. Consistent with the concepts of syntax and semantics, an additional layer can be introduced to capture geometric information, which gives rise to the tensorial structure depicted in Figure~\ref{fig:python_times_geometry}. This structure can be understood as follows: an output is considered acceptable to the user only if it is valid on both the syntax and semantic levels; applied to both the Python code and the corresponding geometry. 

\begin{figure}[h!]
    \centering
    \includegraphics[width=0.7\textwidth]{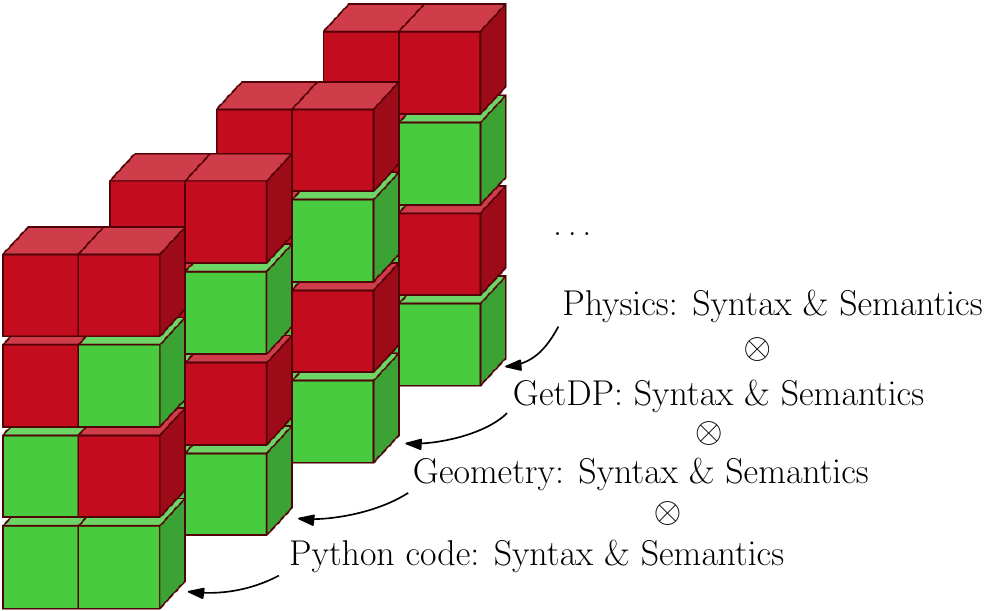}
    \caption{Stack of syntaxes and semantics for workflow as in Section~\ref{subsec:inf_dsl_code_w_examples} and \ref{subsec:inferring_dsl_excluding_examples}.}
    \label{fig:python_times_geometry_times_etc}
\end{figure}

Conceptually, for the workflow presented in Sections~\ref{subsec:inf_dsl_code_w_examples} and \ref{subsec:inferring_dsl_excluding_examples}, two additional layers are introduced (see Figure~\ref{fig:python_times_geometry_times_etc}). Here, also the inferred GetDP code might contain syntax errors, such as improperly matched curly brackets. Semantic errors arise when a field variable is accessed that does not correspond to a primary variable in the FE formulation. Again, these errors are not difficult to detect, as GetDP’s parser will produce an error message.
Again more subtle, the inferred formula for the magnetic energy density~$w_m$ must be physically consistent. In this context, syntactical errors can be interpreted as operations that violate mathematical syntax, such as the addition of scalar and vector fields, consider, e.g., the expression ``$v + \mathbf{A}$''.  Semantic errors, on the other hand, refer to operations that are mathematically valid but physically meaningless, for example, the addition of different vector field quantities such as ``$\mathbf{E} + \mathbf{H}$''.

The dynamically generated textual summary (see Section~\ref{subsec:textual_summary}) introduces yet another layer. In this context, syntactical errors can be interpreted as grammatical or spelling errors in the generated text. Semantic errors, on the other hand, occur when there is a mismatch between the actual generated model and its corresponding textual summary.

\section{Evaluation of the AI workflow}
\label{sec:evaluation}
The goal of this section is to provide some quantifiable results regarding the presented AI workflow that lay the foundation for more in-depth evaluation analysis in future research. 
\subsection{Setup}
\label{subsec:eval_setup}
\begin{figure}[h!]
\centering
\begin{subfigure}[b]{0.32\textwidth}
    \centering
    \includegraphics[width=\textwidth,height=4cm,keepaspectratio]{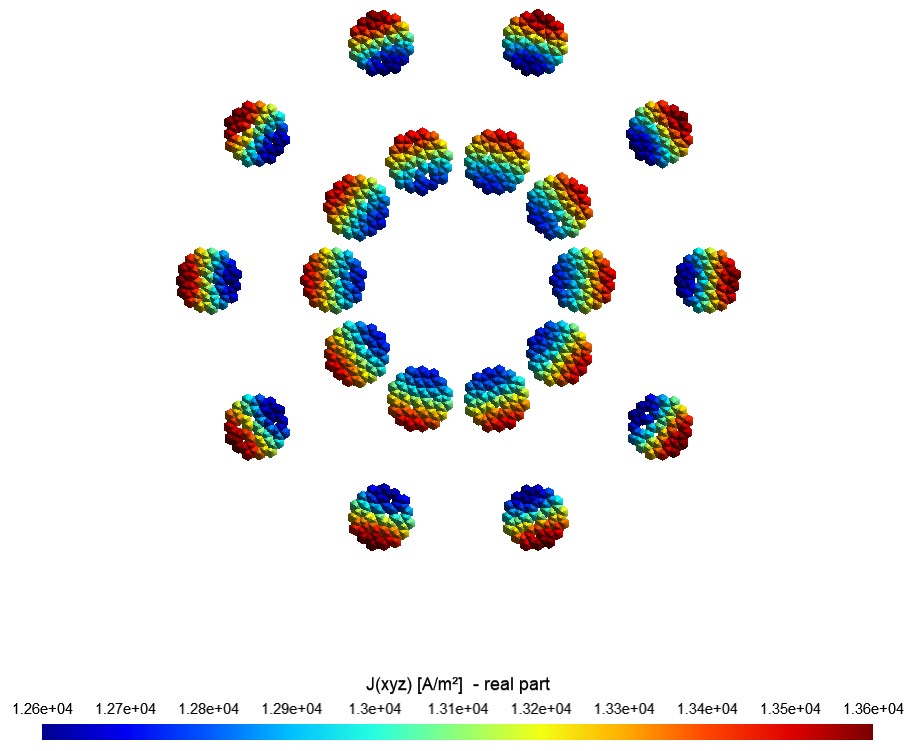}
    \caption{Geometry (basic)}
    \label{fig:sub1}
\end{subfigure}
\hfill
\begin{subfigure}[b]{0.32\textwidth}
    \centering
    \includegraphics[width=\textwidth,height=4cm,keepaspectratio]{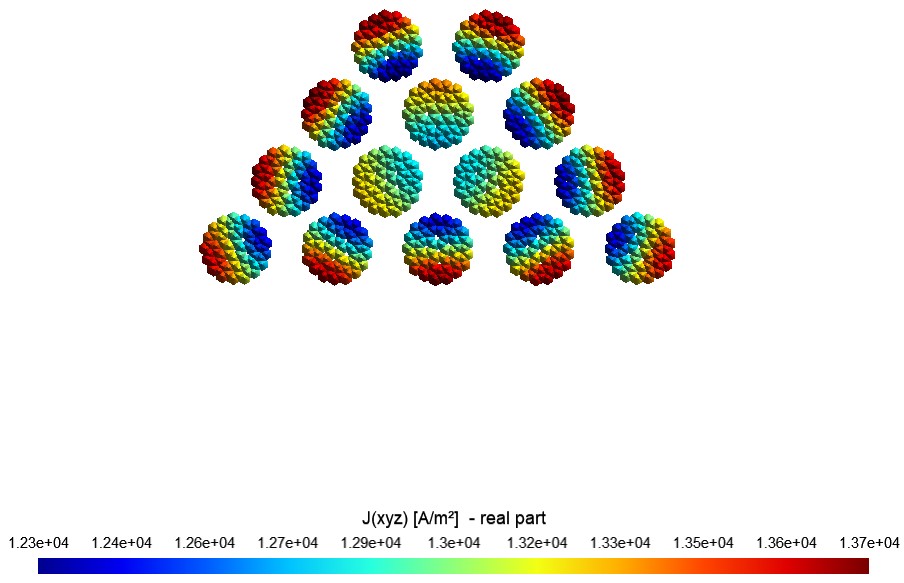}
    \caption{Geometry (intermediate)}
    \label{fig:sub2}
\end{subfigure}
\hfill
\begin{subfigure}[b]{0.32\textwidth}
    \centering
    \includegraphics[width=\textwidth,height=4cm,keepaspectratio]{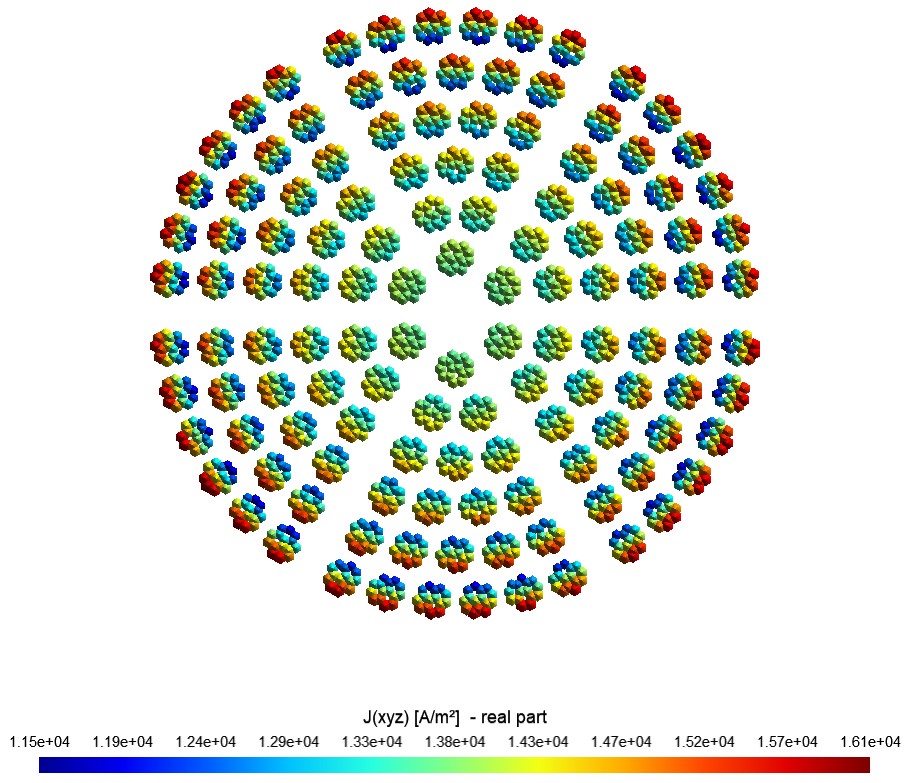}
    \caption{Geometry (advanced)}
    \label{fig:sub3}
\end{subfigure}

\vspace{0.5cm} 

\begin{subfigure}[b]{0.32\textwidth}
    \centering
    \includegraphics[width=\textwidth,height=4cm,keepaspectratio]{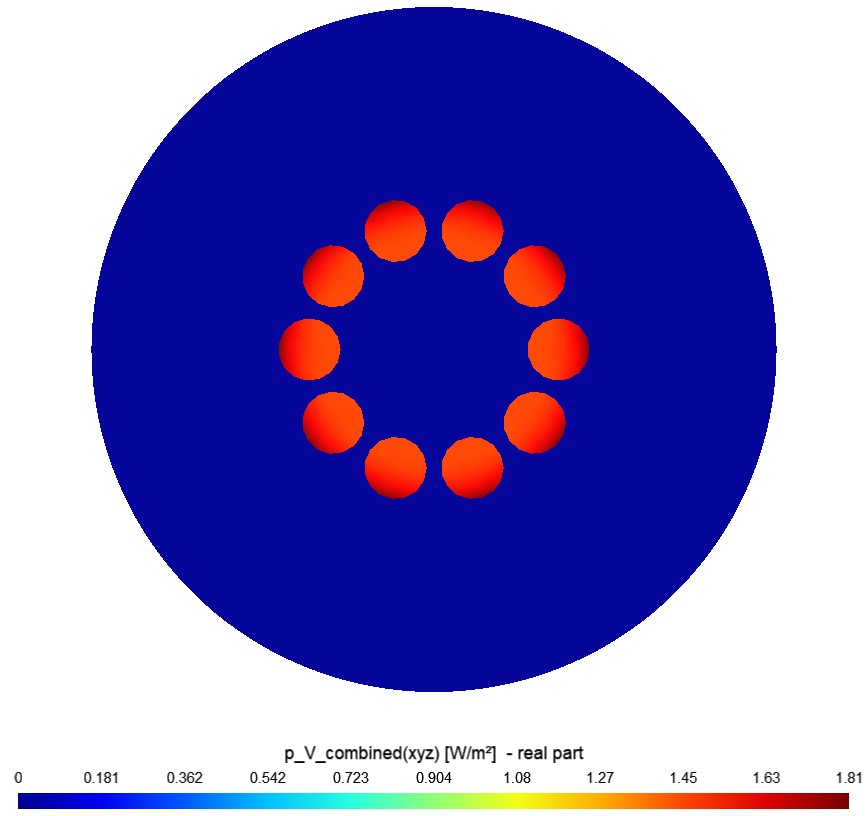}
    \caption{PP (basic)}
    \label{fig:sub4}
\end{subfigure}
\hfill
\begin{subfigure}[b]{0.32\textwidth}
    \centering
    \includegraphics[width=\textwidth,height=4cm,keepaspectratio]{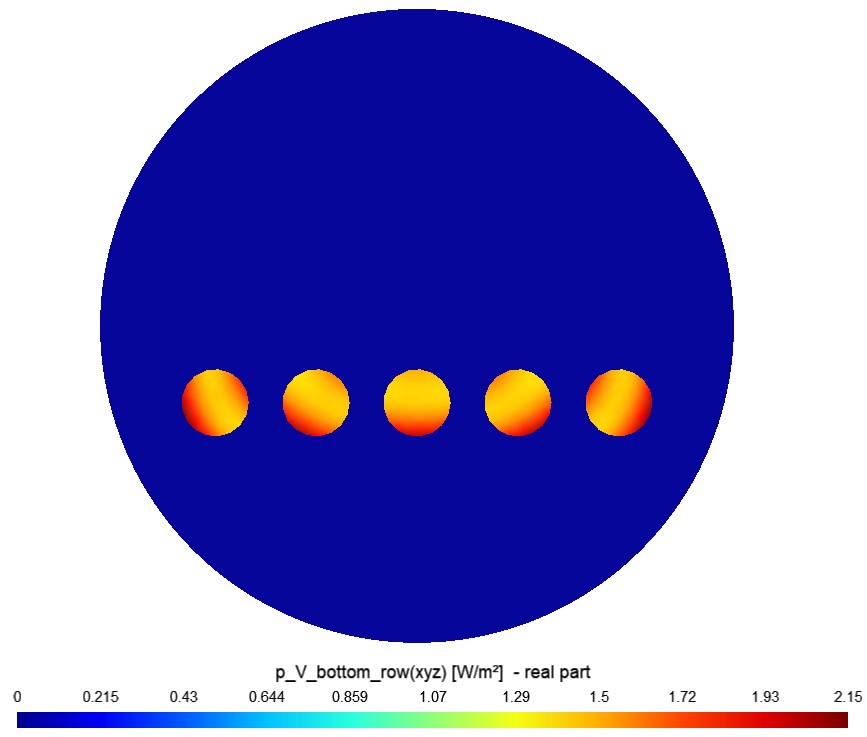}
    \caption{PP (intermediate)}
    \label{fig:sub5}
\end{subfigure}
\hfill
\begin{subfigure}[b]{0.32\textwidth}
    \centering
    \includegraphics[width=\textwidth,height=4cm,keepaspectratio]{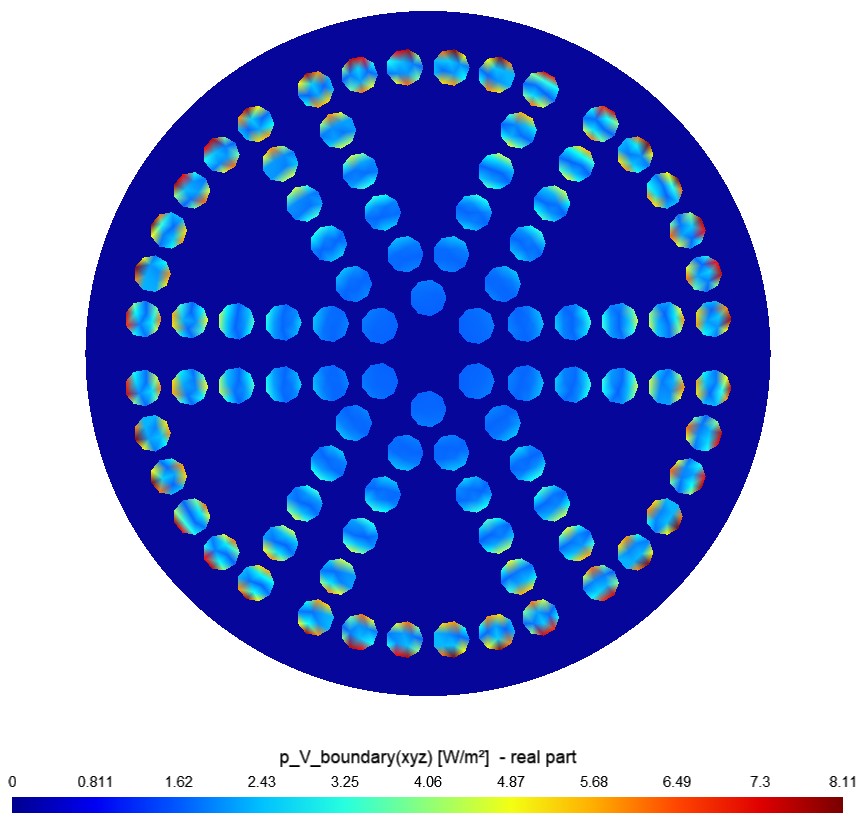}
    \caption{PP (advanced)}
    \label{fig:sub6}
\end{subfigure}

\caption{Reference simulations based on benchmarking user prompts (see Table~\ref{tab:prompts}); ``PP'' denotes post-processing.}
\label{fig:benchmarking_user_prompts_based_reference_simulations}
\end{figure}

Our proposed evaluation setup is as follows: 
\begin{enumerate}
    \item We choose three benchmarking user prompts (see Table~\ref{tab:prompts}) that represent electrical conductors in different geometrical configurations (recall Section~\ref{subsec:inf_pyt_code_gen_list_coord_tuples}) and post-processing routine executions (recall Section~\ref{subsec:inferring_dsl_excluding_examples}). These prompts represent different levels of complexity, i.e., basic, intermediate, and advanced. The complexity of user prompts is assessed across the three tiers based on the ambiguity potential and the electromagnetic computation demands as judged by a subject matter expert. Mind that the system prompts (cf.~Appendix~\ref{sec:syste_prompts}) are kept constant for all runs.
    \item Since the model Gemini-2.0-Flash is expected to be deprecated by publication (cf.~\cite{deprecations}), we select the following five models for the analysis:
         \begin{enumerate}
         \item Gemma-3-1b-It 
         \item Gemma-3-27b-It 
         \item Gemini-2.5-Flash-Lite
         \item Gemini-2.5-Flash 
         \item Gemini-3.1-Fash-Lite 
         \end{enumerate}

   Gemma models are lightweight, open-weight models built on technology similar to that of Gemini's full-fledged, closed-source lineup. Both model families are consumed via Google's corresponding APIs. Note that Gemini-2.5-Flash has higher input and output token prices than Gemini-3.1-Flash-Lite, which in turn exceeds Gemini-2.5-Flash-Lite (cf.~\cite{pricing}). These prices indicate as a first approximation that, for the Gemini models, Gemini-2.5-Flash is the most capable, followed by Gemini-3.1-Flash-Lite, then Gemini-2.5-Flash-Lite.
   \item For a given model and a benchmarking user prompt, the number of tries until the first successful \textit{syntactical} AI workflow execution is counted (recall Section~\ref{subsec:syntax_and_semantics}). Two independent attempts are conducted. If more than ten tries are needed, an ``x'' is marked. The results are shown in Table~\ref{tab:no_tries_first_suc_prog_exec}.
   \item Based on the results from Table~\ref{tab:no_tries_first_suc_prog_exec}, it is decided to proceed with a subset of the models to analyze the successful \textit{semantical} AI workflow execution (recall Section~\ref{subsec:syntax_and_semantics}). In Figure~\ref{fig:benchmarking_user_prompts_based_reference_simulations}, there are reference simulation results that show one representative of the class of valid geometrical configurations and valid post-processing routine executions associated with each benchmarking user prompt. Given these reference simulation results, a subject matter expert evaluates each successful \textit{syntactical} AI workflow execution whether there is (a) a valid geometrical configuration and (b) given (a), whether there is a valid post-processing routine execution in the sense of that they are members of the class of valid geometrical configurations and valid post-processing routine executions, respectively. \\
   For example, for one attempt, if there are eight valid geometrical configurations out of ten successful \textit{syntactical} AI workflow executions, then $8/10$ is recorded. Given the eight valid geometrical configurations, if there are five valid post-processing routine executions, then $5/8$ is recorded. In total two attempts for each pair of models and benchmarking user prompts are conducted. The results are summarized in Table~\ref{tab:successful_semantical_ai_workflow_execution_records}.
\end{enumerate}

\begin{table}[h!]
\centering
\renewcommand{\arraystretch}{1.4}
\setlength{\tabcolsep}{8pt}
\begin{tabular}{p{0.95\textwidth}}
\toprule
\textbf{Basic:} ``Run an MQS simulation with ten conductors placed along an outer circle (with a radius of 4 cm) and ten conductors placed in an inner circle. Note that each conductor has a radius of 5 mm. Ensure that the conductors are not overlapping. Plot the ohmic power loss density only for the inner circle's conductors.'' \\
\midrule
\textbf{Intermediate:} ``Run an MQS simulation model with a proper number of conductors that fill completely a symmetric trapezoidal slot form. Ensure that, at each vertex of the symmetric trapezoidal slot form, there is a conductor. Avoid that the conductor cross sections are overlapping when considering the radius = 0.005 m. Please enlarge the spacing between the conductors. Plot the ohmic power loss density only for the bottom row of conductors.'' \\
\midrule
\textbf{Advanced:} ``Run an MQS simulation with a proper number of conductors that are formed like a Milliken-type conductor. Use at least 100 conductors. Arrange this number of conductors in separated segments that are kind of pie-shaped. Typically you need 6 of them. Leave some space in the inner part of the total cross-section (ca. 0.01 m) such that there is no conductor. Leave also some space between the segments such that their separation is clearly visible. Avoid that the conductor cross sections are overlapping when considering the radius = 0.005 m. Please enlarge the spacing between the conductors. Plot the ohmic power loss density only for the conductors of each segment that are along the boundary of the segment.'' \\
\bottomrule
\end{tabular}
\caption{Benchmarking user prompts (basic, intermediate, advanced) for the evaluation of the AI workflow that are representing electrical conductors in different geometrical configurations (recall Section~\ref{subsec:inf_pyt_code_gen_list_coord_tuples}) and post-processing routines (recall Section~\ref{subsec:inferring_dsl_excluding_examples}).}
\label{tab:prompts}
\end{table}

\definecolor{myLightGray}{gray}{0.8}
\definecolor{myMediumGray}{gray}{0.4}

\begin{table}[!ht]
\centering
\setlength{\tabcolsep}{12pt}
\renewcommand{\arraystretch}{1.3}

\begin{tabular}{l|ccc}
\toprule
 & \textbf{Basic} & \textbf{Intermediate} & \textbf{Advanced} \\
\midrule
\textbf{Gemma-3-1b-It} & {\color{myLightGray}x\;\color{black}\vrule width 1pt\;\color{myMediumGray}x} & {\color{myLightGray}x\;\color{black}\vrule width 1pt\;\color{myMediumGray}x} & {\color{myLightGray}x\;\color{black}\vrule width 1pt\;\color{myMediumGray}x} \\
\textbf{Gemma-3-27b-It} & {\color{myLightGray}1\;\color{black}\vrule width 1pt\;\color{myMediumGray}1} & {\color{myLightGray}1\;\color{black}\vrule width 1pt\;\color{myMediumGray}2} & {\color{myLightGray}x\;\color{black}\vrule width 1pt\;\color{myMediumGray}x} \\
\textbf{Gemini-2.5-Flash-Lite} & {\color{myLightGray}1\;\color{black}\vrule width 1pt\;\color{myMediumGray}2} & {\color{myLightGray}3\;\color{black}\vrule width 1pt\;\color{myMediumGray}2} & {\color{myLightGray}x\;\color{black}\vrule width 1pt\;\color{myMediumGray}9} \\
\textbf{Gemini-2.5-Flash} & {\color{myLightGray}1\;\color{black}\vrule width 1pt\;\color{myMediumGray}1} & {\color{myLightGray}1\;\color{black}\vrule width 1pt\;\color{myMediumGray}1} & {\color{myLightGray}2\;\color{black}\vrule width 1pt\;\color{myMediumGray}5} \\
\textbf{Gemini-3.1-Flash-Lite} & {\color{myLightGray}2\;\color{black}\vrule width 1pt\;\color{myMediumGray}1} & {\color{myLightGray}1\;\color{black}\vrule width 1pt\;\color{myMediumGray}1} & {\color{myLightGray}x\;\color{black}\vrule width 1pt\;\color{myMediumGray}x} \\
\bottomrule
\end{tabular}
\caption{For a given model (Gemma-3-1b-It, Gemma-3-27b-It, Gemini-2.5-Flash-Lite, Gemini-2.5-Flash, and Gemini-3.1-Flash-Lite) and a given benchmarking user prompt (basic, intermediate, advanced), the number of tries until the first successful \textit{syntactical} AI workflow execution is recorded. The light-gray numbers and gray numbers indicate results from the first attempt and the second attempt, respectively. An ``x'' indicates more than ten tries without success.}
\label{tab:no_tries_first_suc_prog_exec}
\end{table}

\definecolor{myLightGray}{gray}{0.8}
\definecolor{myMediumGray}{gray}{0.4}

\begin{table}[h!]
\centering 
\setlength{\tabcolsep}{6pt} 
\renewcommand{\arraystretch}{2.2}
\setlength{\arrayrulewidth}{1pt}
\arrayrulecolor{black}
\resizebox{\textwidth}{!}{
\begin{tabular}{l|ccc}
\toprule
 & \textbf{Basic} & \textbf{Intermediate} & \textbf{Advanced} \\
\midrule
\textbf{Gemma-3-27b-It} & 
  \makecell{
    \begin{tabular}{l|l}
    (a) \color{myLightGray} 10/10 & (a) \color{myMediumGray} 10/10 \\
    \hline
    (b) \color{myLightGray} 0/10 & (b) \color{myMediumGray} 0/10 
    \end{tabular}
  } &
  \makecell{
    \begin{tabular}{l|l}
    (a) \color{myLightGray} 0/10 & (a) \color{myMediumGray} 0/10 \\
    \hline
    (b) \color{myLightGray} 0 & (b) \color{myMediumGray} 0
    \end{tabular}
  } &
  \makecell{\color{black}\rotatebox{45}{\rule{1.2em}{0.15em}}} \\
\midrule
\textbf{Gemini-2.5-Flash} & 
  \makecell{
    \begin{tabular}{l|l}
    (a) \color{myLightGray} 10/10 & (a) \color{myMediumGray} 10/10 \\
    \hline
    (b) \color{myLightGray}10/10 & (b) \color{myMediumGray} 10/10 
    \end{tabular}
  } &
  \makecell{
    \begin{tabular}{l|l}
    (a) \color{myLightGray}5/10 & (a) \color{myMediumGray} 4/10 \\
    \hline
    (b) \color{myLightGray}5/5 &  (b) \color{myMediumGray} 4/4 
    \end{tabular}
  } &
  \makecell{
    \begin{tabular}{l|l}
    (a) \color{myLightGray}8/10 & (a) \color{myMediumGray} 7/10 \\
    \hline
    (b) \color{myLightGray}8/8 & (b) \color{myMediumGray} 7/7
    \end{tabular}
  } \\
\midrule
\textbf{Gemini-3.1-Flash-Lite} & 
  \makecell{
    \begin{tabular}{l|l}
    (a) \color{myLightGray}10/10 & (a) \color{myMediumGray}10/10 \\
    \hline
    (b) \color{myLightGray}7/10 & (b) \color{myMediumGray}9/10 
    \end{tabular}
  } &
  \makecell{
    \begin{tabular}{l|l}
    (a) \color{myLightGray}2/10 & (a) \color{myMediumGray}1/10 \\
    \hline
    (b) \color{myLightGray}1/2 & (b) \color{myMediumGray}1/1 
    \end{tabular}
  } &
  \makecell{\color{black}\rotatebox{45}{\rule{1.2em}{0.15em}}} \\
\bottomrule
\end{tabular}%
} 
\caption{For a given model (Gemma-3-27b-It, Gemini-2.5-Flash, and Gemini-3.1-Flash-Lite) and a given benchmarking user prompt (basic, intermediate, advanced) and given ten successful \textit{syntactical} AI workflow executions, the relative number of (a) valid geometrical configurations and (b) given (a), there is a valid post-processing routine execution. The light-gray numbers and gray numbers indicate results from the first attempt and the second attempt, respectively.}
\label{tab:successful_semantical_ai_workflow_execution_records}
\end{table}


\subsection{Observations in terms of quality}
\label{subsec:observations_eval_setup}

Given the results in Table~\ref{tab:no_tries_first_suc_prog_exec}, using Gemma-3-1b-It, one is not able to receive a successful \textit{syntactical} AI workflow execution for any of the benchmarking user prompts. The advanced benchmarking user prompt seems to be the most challenging for all five models, though, Gemini-2.5-Flash appears to be the only reliable one that, on average, needs 3.5 tries for getting a successful syntactical AI workflow execution for the advanced case. The basic and the intermediate benchmarking user prompt are feasible for most of the models. However, Gemini-2.5-Flash-Lite needs, on average, slightly more tries than the other models. Given these results, we conclude that, in terms of gain of knowledge for the next evaluation step, i.e., the successful \textit{semantical} AI workflow execution analysis, it is sufficient to consider solely the models Gemini-2.5-Flash, Gemini-3.1-Flash-Lite, and Gemma-3-27b-It.

\begin{figure}[h!]
    \centering
    \includegraphics[width=0.3\linewidth]{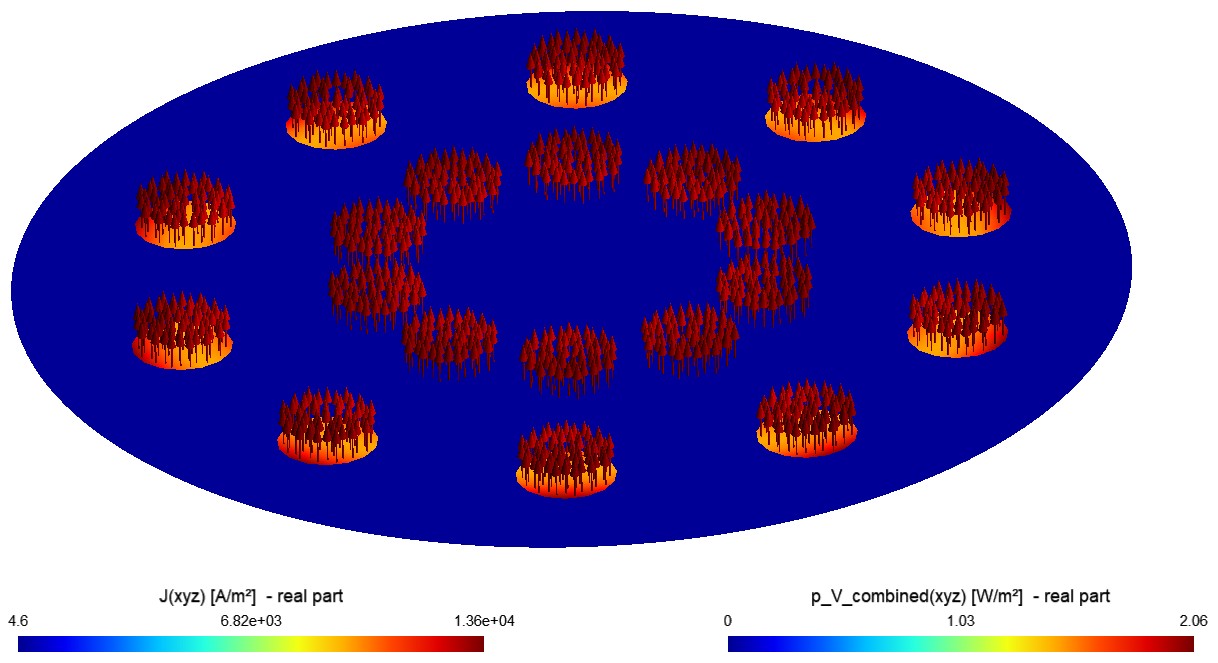}
    \includegraphics[width=0.69\linewidth]{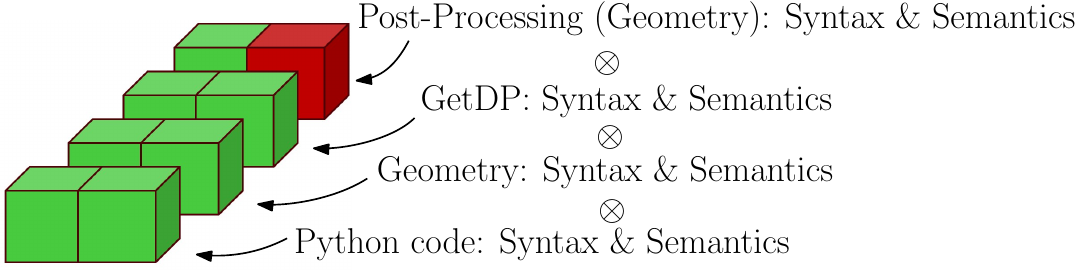}
    \caption{Left: Systems result for the basic benchmarking prompt using the LLM Gemma-3-27b-It. The correct conductor arrangement was achieved, however the ohmic power loss density is wrongly plotted for the outer layer of conductors. Right: Corresponding interpretation of the stack of syntaxes and semantics, cf. discussion in Sec.~\ref{subsec:syntax_and_semantics}.}
    \label{fig:wrong_geometrical_semantics}
\end{figure}

Given the results in Table~\ref{tab:successful_semantical_ai_workflow_execution_records}, it is clear that Gemma-3-27b-It is not able to properly lead to successful \textit{semantical} AI workflow executions for the intermediate benchmarking user prompt, more precisely, there has not been observed one valid geometrical configurations, i.e., a symmetric trapezoidal slot form. In the case of the basic benchmarking user prompt, there has not been observed one valid post-processing routine execution (given a valid geometric configuration), more precisely, the ohmic power loss density has not been plotted for the inner circle's conductors, cf. Figure~\ref{fig:wrong_geometrical_semantics}.

Using Gemini-3.1-Flash-Lite for the basic benchmarking prompt leads, on average, almost surely to a valid geometric configuration and, given those valid geometric configurations, on average, in 80\% of cases, it leads to a valid post-processing routine execution. Using it for the intermediate benchmarking prompt leads, on average, only in 15\% of cases to a valid geometric configuration and, given those valid geometric configurations, on average, in 75\% of cases, it leads to a valid post-processing routine execution. \\
Using Gemini-2.5-Flash for the basic benchmarking prompt leads, on average, almost surely to a valid geometric configuration and, given those valid geometric configurations, on average, almost surely, it leads to a valid post-processing routine execution as well. Using it for the intermediate benchmarking prompt leads, on average, only in 45\% of cases to a valid geometric configuration and, given those valid geometric configurations, on average, almost surely, it leads to a valid post-processing routine execution. Finally, utilizing Gemini-2.5-Flash for the intermediate benchmarking prompt leads, on average, in 75\% of cases to a valid geometric configuration and, given those valid geometric configurations, on average, almost surely, it leads to a valid post-processing routine execution.

Given the above-mentioned observations, it is reasonable to assume that using Gemini-2.5-Flash from Google's AI model families will lead to successful \textit{syntactical} and \textit{semantical} AI workflow executions.

\subsection{Observations in terms of costs and time}
\label{subsec:remark_benefits_time_to_experimentation}

\textbf{Costs}.
The previous Section~\ref{subsec:observations_eval_setup} focused primarily on the quality of the AI workflow. Regarding costs, each benchmarking user prompt (plus constant system prompt; cf. Section~\ref{subsec:eval_setup}) averages approximately 10,000 input and output tokens such that all runs in Tables~\ref{tab:no_tries_first_suc_prog_exec} and~\ref{tab:successful_semantical_ai_workflow_execution_records} total roughly \euro $1$.\\\\
\textbf{Time}.
A single run from the Table~\ref{tab:no_tries_first_suc_prog_exec} and the Table~\ref{tab:successful_semantical_ai_workflow_execution_records} takes just a few seconds. Note the up-front setup investment and ongoing maintenance, but post-setup, time-to-ex\-pe\-ri\-mentation drops by orders of magnitude. Section~\ref{subsec:observations_eval_setup} confirms acceptable results across complexity levels in seconds.\\
Contrast this with simulation engineers without AI tools: For the basic prompt (see Table~\ref{tab:prompts}), an entry-level engineer might need 8 hours, mid-level 4 hours, and master-level 2 hours to start experimenting. Subsequently, for the intermediate prompt, these times halve; and for the advanced one, the times (most likely) increase again — even for master-level simulation engineers. Even with AI tools, our proposed AI workflow likely offers lower time-to-experimentation, reliably and cost-efficiently, across more use cases.



\section{Conclusion and outlook}
\label{sec:conclusions_and_outlook}


\subsection{Conclusion}

This work has investigated an LLM-based chatbot for electromagnetic simulations as a minimum viable setup within the context of AI-assisted code generations and executions for numerical experiments. 

More precisely, we have presented a workflow that is composed by open source FE tools (Gmsh and GetDP) and a large language model (Google Gemini-2.0-Flash) embedded within a common Python-based interface (cf. Section~\ref{sec:workflow}). The basic AI workflow presented in Section~\ref{subsec:inf_pyt_code_gen_list_coord_tuples} infers Python code to generate a list of coordinate tuples. However, architectural extensions also enabled the inference of domain-specific language code, with and without meaningful examples in the system prompt (see Sections~\ref{subsec:inf_dsl_code_w_examples} and \ref{subsec:inferring_dsl_excluding_examples}), and the inference of a textual summary of the AI workflow’s output (see Section~\ref{subsec:textual_summary}). Finally, we have examined the semantic and syntactic sources of potential failure within the AI workflow (see Section~\ref{subsec:syntax_and_semantics}) and quantified these by introducing an evaluation methodology in Section~\ref{sec:evaluation}.

One key insight is that the provided setup can already enable a useful level of automation of a numerical simulation workflow which can significantly reduce the time required to generate a well-posed numerical simulation model. Mind that this time-to-experimentation is a critical metric and lowering it facilitates the accelerated exploration of, for instance, various physical scenarios and corresponding numerical simulation model configurations.

Another key insight is that relying merely on the internal knowledge of an LLM (that is, its training data) as a memory mechanism leads to insufficient outcomes of the AI workflow based on human evaluation. For example, physically relevant factors are not taken into account properly. However, by additionally using the LLM's context (via the user prompt and the system prompt) as a further memory mechanism satisfactory outcomes of the AI workflow can be achieved.

A third key insight is that the systematic consideration of both the semantic and syntactic aspects of an AI workflow's essential elements can offer a valuable conceptual guidance for analyzing potential failure modes. The proposed visual representation of a stack of syntaxes and semantics for the workflow's essential elements illustrates the many combinations in which the workflow can fail. Note that this proposed visual representation is highly scalable with respect to the increasing complexity of the underlying setup, thus offering a way to conceptually manage the increased complexity.

A fourth key insight is that the shown AI workflow facilitates a declarative development style where the focus is on describing what the desired outcome is, rather than explicitly specifying how to achieve the outcome step-by-step. In many scenarios, the desired outcome is a correct numerical solution, regardless of the exact method used to achieve it. However, for validating the solution's correctness, domain knowledge is needed -- which leads to the last key insight. 

The final key insight is that, due to the probabilistic nature of an LLM, it is unclear how to construct a reliable automated evaluation method for the AI workflow. The evaluation of the discussed AI workflow's outcomes relied on human evaluation. Human evaluation defines one end of the spectrum of evaluation methods, formal verification defines the other end of the spectrum. Since most likely no formal guarantees regarding the AI workflow's outcomes can be given, it seems at least conceivable to design semi-automated evaluation methods where human evaluation is potentially reduced to a minimum. An in-depth analysis of these ideas is left for future investigations.


\subsection{Outlook}

Future research and development efforts can build on the presented results in several directions. One of the most challenging and crucial efforts is the investigation of reliable (semi-) automated evaluation methods for the AI workflow. For evaluating the quality of textual summaries, for instance, one promising approach could be the usage of an additional LLM that is adapted for evaluation purposes. Another promising approach involves the use of embeddings, or vector representations \cite{weller2025theoreticallimitationsembeddingbasedretrieval}, along with similarity metrics that enable a textual summary to be quantitatively compared to human-curated or AI-generated benchmarks. 

Addressing (semi-)automated evaluation methods is also important for systematically examining the space spanned by system prompts and LLMs in order to find the optimal configuration for the given AI workflow.

Furthermore, examining (semi-) automated evaluation methods is also critical in the investigation of more complex AI workflows. Notice that, from a control-flow graph point of view, the discussed AI workflow can be mathematically represented as a directed acyclic graph. However, more intricate graphs are conceivable that include both sequential and parallel paths as well as iterations and selections. Moreover, if more tools (i.e., if more callable functions) are available and the planning (i.e., the construction of a control-flow graph) to complete a user-defined task is undertaken by an LLM, then the resulting system is more appropriately characterized as an AI agent rather than a predefined AI~workflow. It should be noted that the research field of AI agents currently lacks a well-defined theoretical foundation (see, e.g., \cite[pp.~276]{Huyen2025}). Therefore, there are many open research questions on AI agents in the context of electromagnetic simulations that deserve a thorough investigation.

Another potential architectural extension to further enhance the developed chatbot’s functionalities is the use of retrieval-augmented generation (RAG) techniques (see, e.g., \cite[pp.~253-275]{Huyen2025}). These techniques would enable the underlying LLM to access and integrate information that the model was not, or only partially, trained on. Additionally, this information is more relevant to a given user prompt than a predefined and user-prompt-agnostic system prompt. In particular, with the FE software tools (Gmsh and GetDP) integrated in the chatbot, it would be beneficial to provide to the LLM only those portions of code repositories or manuals of the FE software tools that are most relevant to a user prompt.

Lastly, a valuable opportunity for future work involves expanding the set of open source numerical tools beyond Gmsh and GetDP to include additional frameworks such as openCFS~\cite{schoder2025opencfsopensourcefinite} and DeepXDE~\cite{DeepXDE}. Such an extension would enable a unified platform of diverse numerical simulation software that can be accessed by user prompts in natural language.


\clearpage
\appendix


\section{System prompts}
\label{sec:syste_prompts}

This appendix includes representative code examples from the system prompts of the workflows discussed in Section~\ref{sec:Architectural_extensions_and_case_studies}. Each code snippet is part of a Python multi-line string that is passed to the LLM. Mind that the full system prompt that is containing a task description, some rules and examples as well as the user input placeholder is only provided for the first system prompt~\ref{subsec:system_prompts_Python}. For the other system prompts, only some examples are shown and, for the sake of brevity, the rest is omitted.

Note that in the system prompts, curly braces ``\{'' and ``\}'' serve a special function, as they are used to denote placeholders for dynamic content. Therefore, when literal braces are required in the string (i.e., not as placeholders), they must be escaped by doubling them — written as ``\{\{'' and ``\}\}''. This ensures that the braces are interpreted as characters rather than as variable placeholders during template rendering. This is particularly necessary for the GetDP syntax code.

\subsection{System prompt for inferring Python code to generate a list of coordinate tuples}
\label{subsec:system_prompts_Python}
\begin{lstlisting}
"""
Task:
You are an expert Python programmer. Your task is to generate Python code that 
creates a list of 2D points based on user specifications. 
Write clean Python code that defines these points as a list of two-element float tuples.

Rules:
- Use only the provided libraries and functions: numpy, matplotlib.pyplot, random. Do not use any other 
libraries or functions that are not explicitly provided.
- Do not include any explanations or apologies in your responses.
- Ensure the function is syntactically correct and uses valid Python methods.
- Call the generated function afterwards.
- Don't make up any result, i.e., if you don't know how to implement a specific feature, 
say you don't know.
- Interpret the variable xy_list as type List[Tuple[float, float]]
- Do not produce markdown.
- Produce only a valid python string-expression.
        
Example 1: "Run a minimal magnetoquasistatic simulation with some initial points"
          from typing import List, Tuple
          from simulation_functions.test_gmsh_api_given_list import run_simulation_from_terminal
          xy_list : List[Tuple[float, float]] = [(0.0, 0.0), (0.02, 0.0), (0.0, 0.02)]
          run_simulation_from_terminal(xy_list)
        
Example 2: "Run an initial mqs simulation using only one conductor"
          from typing import List, Tuple
          from simulation_functions.test_gmsh_api_given_list import run_simulation_from_terminal
          xy_list : List[Tuple[float, float]] = [(0.0, 0.0)]
          run_simulation_from_terminal(xy_list)

(...)

The user input is:
{user_input}
"""
\end{lstlisting}

\clearpage
\subsection{System prompt for inferring domain-specific language code (including meaningful examples in the system prompt)}
\label{subsec:system_prompts_including_examples}
\begin{lstlisting}
Example 6: "Evaluate the plot of the power loss density for the first conductor of three conductors along the x-axis"

  with open("simulation_functions/round_2D_conductors_extra_PP.pro", "w") as file:
    text_1 = "PostProcessing {{
      {{ Name MagDyn_b; NameOfFormulation MagDyn_a; 
        PostQuantity {{
          {{ Name OhmicLossDensity_conductor_1; 
           Value {{ Local {{ [sigma[]/2 * Norm[ (- Dt[{{a}}] - {{grad_phi}})]^2 ]; 
           In Region[{{Omega_c_1}}]; Jacobian Vol; }} }} }} 
          }} // end of PostQuantity block
       }} // end of specific PostProcessing MagDyn_b
    }} // end of PostProcessing block
    "                    
    file.write(text_1)

  with open("simulation_functions/round_2D_conductors_extra_PO.pro", "w") as file:
    text_1 = "PostOperation {{
      {{ Name MagDyn_b; NameOfPostProcessing MagDyn_b; 
        Operation {{
            Print[ OhmicLossDensity_conductor_1, OnElementsOf Omega, 
                   File "Results/p_V_conductor_selected.pos", 
                   Name "p_V_c_1(xyz) [W/m^3] ", Format Gmsh 
                 ];
        }} // end of Operation block
      }} // end of specific PostOperation MagDyn_b
   }} // end of PostOperation block
   "                    
   file.write(text_1)
        
  from typing import List, Tuple
  from simulation_functions.test_gmsh_api_given_list import run_simulation_from_terminal
  xy_list : List[Tuple[float, float]] = [(0.02, 0.0), (0.0, 0.0), (-0.02, 0.0)]
  run_simulation_from_terminal(xy_list)
\end{lstlisting}

\subsection{System prompt for inferring domain-specific language code (excluding meaningful examples in the system prompt)}
\label{subsec:system_prompts_excluding_examples}
\begin{lstlisting}
Example 12: "Evaluate the plot of the magnetic vector field of three conductors along the x-axis"

  with open("simulation_functions/round_2D_conductors_extra_PP_inference.pro", "w") as file:
    text_1 = "PostProcessing {{
      {{ Name MagDyn_c; NameOfFormulation MagDyn_a; 
        PostQuantity {{
          {{ Name h_vector_field; 
           Value {{ Term {{ [nu[] * {{d a}}]; 
           In Omega; Jacobian Vol; }} }} }} 
          }} // end of PostQuantity block
       }} // end of specific PostProcessing MagDyn_b
    }} // end of PostProcessing block
    "                    
    file.write(text_1)

  with open("simulation_functions/round_2D_conductors_extra_PO_inference.pro", "w") as file:
    text_1 = "PostOperation {{
      {{ Name MagDyn_c; NameOfPostProcessing MagDyn_c; 
        Operation {{
            Print[ h_vector_field, OnElementsOf Omega, 
                   File "Results/h_vector_field.pos", 
                   Name "H(xyz) [A/m] ", Format Gmsh 
                 ];
        }} // end of Operation block
      }} // end of specific PostOperation MagDyn_c
   }} // end of PostOperation block
   "                    
   file.write(text_1)
        
  from typing import List, Tuple
  from simulation_functions.test_gmsh_api_given_list import run_simulation_from_terminal
  xy_list : List[Tuple[float, float]] = [(0.02, 0.0), (0.0, 0.0), (-0.02, 0.0)]
  run_simulation_from_terminal(xy_list)
\end{lstlisting}

\subsection{System prompt for inferring a textual summary of the LLM's output}
\label{subsec:system_prompts_textual_summary}
\begin{lstlisting}
Example 4: 
  "Run a basic mqs simulation using three conductors where one is on the y-axis and the other two are on the x-axis"
  
  Plot Execution Result:
  import os
  current_dir = os.getcwd()
  print(current_dir)
  from typing import List, Tuple
  from simulation_functions.test_gmsh_api_given_list import run_simulation_from_terminal
  xy_list : List[Tuple[float, float]] = [(0.02, 0.0), (0.0, 0.02), (-0.02, 0.0)]
  run_simulation_from_terminal(xy_list)

  The output of Example 4 shows a 2D-plot of three electric conductors whose center points are at (0.02, 0.0), (0.0, 0.02), and (-0.02, 0.0) and all radii are 5 mm. Thus, two conductors lie on the x-axis with respect to a Cartesian coordinate frame and one lies on the y-axis. Current density vector fields are observed only within the conductors, in the surrounding non-conducting computational domain, there current density vector fields are zero. All three current density vector fields possess solely a non-zero z-component and are oriented in positive z-direction. Since there are more than one conductor, in addition to the skin effect, the so-call proximity physical effect can be observed. This leads to non-uniformly spatially distributed current density vector fields. The given spatial configuration of the conductors and the spatial direction of the currents results in shift of the current density distribution towards the outer regions of cross-sections of the outer conductors. Due the given spatial configuration of the conductors and the spatial direction of the currents results, the magnetic flux densities associtated with each individual conductor are exhibiting different directions within the computational domain between both conductors, and are exhibiting same directions within the computational domain outside of the in-between area of both conductors. Hence, the total magnetix flux density's magnitude is lower within the computational domain between both conductors, and higher elsewhere. Finally, a shift of the current density distribution towards the outer regions of cross-sections of the conductors can be observed.    
\end{lstlisting}


\clearpage
\section{LLM outputs}
\label{sec:llm_outputs}

This appendix includes representative code examples generated by the LLM. Note that the LLM’s output is originally a multi-line string. Here, we show its cleaned version, i.e., the version that is actually executed by Python (\ref{subsec:llm_outputs_inferred_python_code}) and GetDP (\ref{subsec:llm_outputs_inferred_dsl_including_examples} and \ref{subsec:llm_outputs_inferred_dsl_excluding_examples}).

\subsection{LLM outputs for inferred Python code to generate a list of coordinate tuples}
\label{subsec:llm_outputs_inferred_python_code}
The Python code below generates the conductor arrangement shown in Figure~\ref{fig:results_inferring_python_code_1} (as a consequence of user prompt~(c)).
\begin{lstlisting}
import numpy as np
from typing import List, Tuple
import matplotlib.pyplot as plt

def generate_hexagonal_grid_points(num_x: int, num_y: int, spacing: float) -> List[Tuple[float, float]]:
    """Generates a list of 2D points arranged in a hexagonal grid.

    Args:
        num_x: The number of points in the x-direction.
        num_y: The number of points in the y-direction.
        spacing: The spacing between the points.

    Returns:
        A list of tuples, where each tuple is a 2D point (x, y).
    """
    points: List[Tuple[float, float]] = []
    for i in range(num_x):
        for j in range(num_y):
            x = i * spacing
            y = j * spacing
            if i % 2 == 1:
                y += spacing / 2
            points.append((x, y))
    return points


from simulation_functions.test_gmsh_api_given_list import run_simulation_from_terminal
num_x = 10
num_y = 10
spacing = 0.02
xy_list: List[Tuple[float, float]] = generate_hexagonal_grid_points(num_x, num_y, spacing)
\end{lstlisting}

\subsection{LLM outputs for inferred domain-specific language code (including meaningful examples in the system prompt)}
\label{subsec:llm_outputs_inferred_dsl_including_examples}
The GetDP code below generates the custom post-processing routine for the ohmic power loss density~$p_\Omega$ shown in Figure~\ref{fig:results_inferring_dsl_code} (as a consequence
of user prompt (i)).

\begin{lstlisting}
PostProcessing{
    { Name MagDyn_b ; NameOfFormulation MagDyn_a ;
        PostQuantity {
                { Name OhmicLossDensity_conductor_4 ; Value { Local { [ sigma[]/2 * Norm[ (- Dt[{a}] - {grad_phi}) ]^2 ] ; In Region[{Omega_c_4}] ; Jacobian Vol ; } } }
        }
    }
}

PostOperation{
    { Name MagDyn_b ; NameOfPostProcessing MagDyn_b ;
        Operation {
                Print[ OhmicLossDensity_conductor_4       , OnElementsOf Omega , File "Results/p_V_conductor_selected.pos" , Name "p_V_c_4(xyz) [W/m^3] "  , Format Gmsh ];
        }
    }
}
\end{lstlisting}

\subsection{LLM outputs for inferred domain-specific language code (excluding meaningful examples in the system prompt)}
\label{subsec:llm_outputs_inferred_dsl_excluding_examples}
The GetDP code below generates the custom post-processing routine for the magnetic energy density~$w_m$ shown on the right hand side of Figure~\ref{fig:workflow_inferring_dsl_code_without_examples} (as a consequence of the user prompt mentioned in Section~\ref{subsec:inferring_dsl_excluding_examples}).

\begin{lstlisting}
PostProcessing{
    {
        Name MagDyn_c;
        NameOfFormulation MagDyn_a;
        PostQuantity {
            {
                Name MagneticEnergyDensity_Diagonal;
                Value {
                    Local {
                        [
                            0.25 * nu[] * Norm[{d a}]^2
                        ];
                        In Region[{Omega_c_1, Omega_c_5, Omega_c_9}];
                        Jacobian Vol;
                    }
                }
            }
        }
    }
}

PostOperation{
    {
        Name MagDyn_c;
        NameOfPostProcessing MagDyn_c;
        Operation {
            Print[
                MagneticEnergyDensity_Diagonal,
                OnElementsOf Omega,
                File "Results/magnetic_energy_density_diagonal.pos",
                Name "w_m_diagonal(xyz) [J/m^3]",
                Format Gmsh
            ];
        }
    }
}
\end{lstlisting}

\clearpage
\printbibliography

@ARTICLE{GetDP,
  author={Dular, P. and Geuzaine, C. and Henrotte, F. and Legros, W.},
  journal={IEEE Transactions on Magnetics}, 
  title={A general environment for the treatment of discrete problems and its application to the finite element method}, 
  year={1998},
  volume={34},
  number={5},
  pages={3395-3398},
  doi={10.1109/20.717799}}

@article{Gmsh,
author = {{Geuzaine, C.} and {Remacle, J.-F.}},
title = {Gmsh: A 3-D finite element mesh generator with built-in pre- and post-processing facilities},
journal = {International Journal for Numerical Methods in Engineering},
volume = {79},
number = {11},
pages = {1309-1331},
doi = {https://doi.org/10.1002/nme.2579},
year = {2009},
}

@ARTICLE{Pinn_Guo,
  author={Guo, Z. and {Nguyen, B.} and {Sabariego, R. V.}},
  journal={IEEE Transactions on Magnetics}, 
  title={Physics-Informed Neural Network for Solving 1-D Nonlinear Time-Domain Magneto-Quasi-Static Problems}, 
  year={2025},
  volume={61},
  number={5},
  pages={1-9},
  doi={10.1109/TMAG.2025.3553236}}

@phdthesis{j_dular_diss,
	AUTHOR = {Dular, J.},
	TITLE = {Standard and Mixed Finite Element Formulations for Systems with Type-II Superconductors},
	YEAR = {2023},
	SCHOOL = {University of Liège, Liege, Belgium},
	SIZE = {239},
	ORGANIZATION = {F.R.S.-FNRS - Fonds de la Recherche Scientifique}
}

@misc{streamlit2025,
  title        = {Streamlit: Build apps for data science},
  author       = {Streamlit},
  howpublished = {\url{https://streamlit.io}},
  year         = {2025},
  note         = {Version 1.41.1, accessed 20 October 2025}
}

@misc{google_gemini_flash_2,
  author       = {{Google DeepMind}},
  title        = {Gemini 2.0 Flash},
  howpublished = {\url{https://ai.google.dev/gemini-api/docs}},
  year         = {2025},
  note         = {Version 2.0, accessed 20 October 2025}
}

@misc{open_foam,
  author       = {{OpenFOAM Ltd}},
  title        = {OpenFOAM},
  howpublished = {\url{https://www.openfoam.com/}},
  year         = {2025},
  note         = {accessed 21 October 2025}
}

@article{neural_operators,
author = {{Kovachki, N.} and {Li, Z.} and {Liu, B.} and {Azizzadenesheli, K.} and {Bhattacharya, K.} and {Stuart, A.} and {Anandkumar, A.}},
title = {Neural operator: learning maps between function spaces with applications to PDEs},
year = {2023},
issue_date = {January 2023},
publisher = {JMLR.org},
volume = {24},
number = {1},
issn = {1532-4435},
journal = {J. Mach. Learn. Res.},
month = jan,
articleno = {89},
numpages = {97}
}

@article{pinn_rodrigo,
    author ={{Rezende, R. S.} and {Schuhmann, R.}},
    title = {An Efficient PINNs Approach Using Hard
Constraints Boundary Conditions for Solving
Electromagnetic Problems},
    journal = {URSI Radio Science Letters},
    volume = {7},
    pages = {1-5},
    year = {2025}, 
    doi = {10.46620/25-0032}
}

@misc{yue2025foamagentautomatedintelligentcfd,
      title={Foam-Agent: Towards Automated Intelligent CFD Workflows}, 
      author={{Yue, L.} and {Somasekharan, N.} and {Cao, Y.} and {Pan, S.}},
      year={2025},
      eprint={2505.04997},
      archivePrefix={arXiv},
      primaryClass={cs.AI},
      url={https://arxiv.org/abs/2505.04997}, 
}

@misc{python,
  author       = {{Python Software Foundation}},
  title        = {Python 3.11.14 documentation},
  year         = {2025},
  url          = {https://docs.python.org/3.11/},
  note         = {Version 3.11.14}
}

@book{Huyen2022,
    author = {Huyen, C.},
    title = {Designing Machine Learning Systems: An Iterative Process for Production-Ready Applications},
    publisher = {{O'Reilly}},
    year = {2022}
}

@book{Huyen2025,
    author = {Huyen, C.},
    title = {AI Engineering: Building Applications with Foundation Models},
    publisher = {{O'Reilly}},
    year = {2025}
}

@misc{weller2025theoreticallimitationsembeddingbasedretrieval,
      title={On the Theoretical Limitations of Embedding-Based Retrieval}, 
      author={{Weller, 0.} and {Boratko, M.} and {Naim, I.} and {Lee, J.}},
      year={2025},
      eprint={2508.21038},
      archivePrefix={arXiv},
      primaryClass={cs.IR},
      url={https://arxiv.org/abs/2508.21038}, 
}

@misc{schoder2025opencfsopensourcefinite,
      title={openCFS: Open Source Finite Element Software for Coupled Field Simulation -- Part Acoustics}, 
      author={{Schoder, S.} and {Roppert, K.}},
      year={2025},
      eprint={2207.04443},
      archivePrefix={arXiv},
      primaryClass={math.NA},
      url={https://arxiv.org/abs/2207.04443}, 
}

@article{DeepXDE,
author = {Lu, L. and Meng, X. and Mao, Z. and Karniadakis, G. E.},
title = {DeepXDE: A Deep Learning Library for Solving Differential Equations},
journal = {SIAM Review},
volume = {63},
number = {1},
pages = {208-228},
year = {2021},
doi = {10.1137/19M1274067},
URL = { 
        https://doi.org/10.1137/19M1274067
},
eprint = { 
        https://doi.org/10.1137/19M1274067
}
}

@article{LuNLoperators,
    author = {{Lu, L.} and {Jin, P.} and {Pang, G.}},
    title = {Learning nonlinear operators via DeepONet based on the universal approximation theorem of operators},
    journal = {Nature Machine Intelligence},
    number = {3},
    pages = {218-219},
    year = {2021},
    doi = {10.1038/s42256-021-00302-5}
}

@article{fourierno,
  author       = {{Li, Z.} and
                  {Kovachki, N.} and
                  {Azizzadenesheli, K.} and 
                  {Liu, B.}},
  title        = {Fourier Neural Operator for Parametric Partial Differential Equations},
  journal      = {CoRR},
  volume       = {abs/2010.08895},
  year         = {2020},
  url          = {https://arxiv.org/abs/2010.08895}
}

@article{raissi1,
title = {Hidden physics models: Machine learning of nonlinear partial differential equations},
journal = {Journal of Computational Physics},
volume = {357},
pages = {125-141},
year = {2018},
issn = {0021-9991},
doi = {10.1016/j.jcp.2017.11.039},
author = {{Raissi, M.} and {Karniadakis, G. E.}}
}

@article{raissi2,
title = {Physics-informed neural networks: A deep learning framework for solving forward and inverse problems involving nonlinear partial differential equations},
journal = {Journal of Computational Physics},
volume = {378},
pages = {686-707},
year = {2019},
issn = {0021-9991},
doi = {https://doi.org/10.1016/j.jcp.2018.10.045},
url = {https://www.sciencedirect.com/science/article/pii/S0021999118307125},
author = {M. Raissi and P. Perdikaris and G.E. Karniadakis}
}

@article{Karniadakis,
    author = {{Karniadakis, G. E.} and {Kevrekidis, I. G.} and {Lu, L.}},
    title = {Physics-informed machine learning},
    journal = {Nature Reviews Physics},
    year = {2021},
    doi = {10.1038/s42254-021-00314-5},
    number = {3},
    pages = {422–440}
}

@article{pinnmaxwell,
    author = {{Lim, J.} and {Psaltis, D.}},
    title = {MaxwellNet: Physics-driven deep neural network training based on Maxwell’s equations},
    journal = {APL Photonics},
    volume = {7},
    number = {1},
    pages = {011301},
    year = {2022},
    issn = {2378-0967},
    doi = {10.1063/5.0071616}
}

@article{ROM,
title = {Model order reduction for parameterized electromagnetic problems using matrix decomposition and deep neural networks},
journal = {Journal of Computational and Applied Mathematics},
volume = {431},
pages = {115271},
year = {2023},
issn = {0377-0427},
doi = {10.1016/j.cam.2023.115271},
author = {{He, X.-F.} and {Li, L.} and {Lanteri, S.} and {Li, K.}}
}

@article{CEMagent,
author = {{Lupoiu, R.} and {Shao, Y.} and {Dai, T.} and {Mao, C.} and {Edée, K.} and {Fan, J. A.} },
title = {A multi-agentic framework for real-time, autonomous freeform metasurface design},
journal = {Science Advances},
volume = {11},
number = {44},
pages = {eadx8006},
year = {2025},
doi = {10.1126/sciadv.adx8006}}

@book{LLMbook,
    author = {{Alammar, J.} and {Grootendorst, M.}},
    title = {Hands-On Large Language Models: Language Understanding and Generation},
    publisher = {{O'Reilly}},
    year = {2024}
}

@misc{deprecations,
  author       = {{Google DeepMind}},
  title        = {Gemini deprecations},
  howpublished = {\url{https://ai.google.dev/gemini-api/docs/deprecations?hl=en}},
  year         = {2026},
  note         = {accessed 08 March 2026}
}

@misc{pricing,
  author       = {{Google DeepMind}},
  title        = {Gemini Developer API pricing},
  howpublished = {\url{https://ai.google.dev/gemini-api/docs/pricing?hl=en}},
  year         = {2026},
  note         = {accessed 08 March 2026}
}


\end{document}